\documentclass[fleqn,usenatbib]{mnras}

\usepackage{newtxtext,newtxmath}
\usepackage{comment}
\usepackage{longtable}
\usepackage{tablefootnote}
\usepackage{threeparttablex}
\usepackage{graphicx}
\usepackage{txfonts}
\usepackage{verbatim}
\usepackage{hyperref}
\usepackage{tabularx}
\usepackage[dvipsnames]{xcolor}
\newcommand{\lsun}{$\log$L/$L_{\odot}\,$}
\newcommand{\msun}{$M$/$M_{\odot}\,$}

\usepackage{multicol}
\usepackage{graphicx}
\usepackage{natbib}
\defcitealias{Desomma2020}{DS20}
\defcitealias{Fiorentino2007}{Fiorentino2007}
\defcitealias{Caputo2005}{Caputo2005}
\defcitealias{Marconi2013}{Marconi2013}
\defcitealias{Wood2006}{Wood2006}

\usepackage[bottom]{footmisc}
\graphicspath{{./}{figures/}}
\usepackage[T1]{fontenc}

\DeclareRobustCommand{\VAN}[3]{#2}
\let\VANthebibliography\thebibliography
\def\thebibliography{\DeclareRobustCommand{\VAN}[3]{##3}\VANthebibliography}

\usepackage{graphicx}	
\usepackage{amsmath}	
\usepackage{amssymb}	

\title[The Cepheid Period-Age and Period-Age-Color relations]{Updated theoretical Period-Age and Period-Age-Color relations for Galactic Classical Cepheids: an application to the \textsl{Gaia} DR2 sample}

\author[G. De Somma et al.]{
Giulia De Somma,$^{1}$$^{2}$$^{3}$\thanks{E-mail: giulia.desomma@inaf.it (University of Naples 'Federico II')}
Marcella Marconi,$^{2}$
Santi Cassisi,$^{4}$$^{5}$ 
Vincenzo Ripepi,$^{2}$
\newauthor{Silvio Leccia$^{2}$},
Roberto Molinaro$^{2}$
and Ilaria Musella$^{2}$
\\
$^{1}$ Dipartimento di Fisica "E. Pancini", Universit\'a di Napoli "Federico II", Compl. Univ. di Monte S. Angelo, Edificio G, Via Cinthia, 80126 Napoli, Italy\\
$^{2}$ INAF-Osservatorio Astronomico di Capodimonte, Via Moiariello 16, 80131 Napoli, Italy\\
$^{3}$ Istituto Nazionale di Fisica Nucleare (INFN)-Sez. di Napoli, Compl. Univ.di Monte S. Angelo, Edificio G, Via Cinthia, 80126 Napoli, Italy\\
$^{4}$ INAF-Osservatorio Astronomico d'Abruzzo, Via Maggini sn, 64100 Teramo, Italy\\
$^{5}$ Istituto Nazionale di Fisica Nucleare (INFN) - Sezione di Pisa, Universit\'a di Pisa, Largo Pontecorvo 3, 56127 Pisa, Italy
}

\date{Accepted 2020 June 22. Received 2020 June 22; in original form 2020 May 12}

\pubyear{2020}

\begin{document}
\label{firstpage}
\pagerange{\pageref{firstpage}--\pageref{lastpage}}
\maketitle

\begin{abstract}
Updated evolutionary and pulsational model predictions are combined in
order to interpret the properties of Galactic Classical Cepheids in the \textsl{Gaia} Data Release 2.
In particular, the location of the instability strip boundaries and the analytical relations connecting pulsation periods to the intrinsic stellar parameters are combined with evolutionary tracks to derive reliable and accurate period-age, and the first theoretical period-age-color relations in the Gaia bands for a solar chemical abundance pattern ($Z$=$0.02$, $Y$=$0.28$). The adopted theoretical framework takes into account possible variations in the mass-luminosity relation for the core helium-burning stage as due to changes in the core convective overshooting and/or mass loss efficiency, as well as the impact on the instability strip boundaries due to different assumptions for superadiabatic convection efficiency.
The inferred period-age and period-age-color relations are applied to a selected sample of both fundamental and first overtone \textsl{Gaia} Cepheids, and individual ages for the various adopted theoretical scenarios are derived.
The retrieved age distributions confirm that a variation in the efficiency of superadiabatic convection in the pulsational model computations has a negligible effect, whereas a brighter Mass-Luminosity relation, as produced by mild overshooting, rotation or mass loss, implies significantly older age predictions. Moreover, older Cepheids are found at larger Galactocentric distances, while first overtone Cepheids are found to be systematically older than the fundamental ones. The comparison with independent age distribution analysis in literature supports the predictive capability of current theoretical framework.

\end{abstract}



\section{Introduction}
Classical Cepheids (CC) are the most important primary distance indicators
in the Local Group and excellent tracers of relatively young (from a few tens to a few hundreds of Myr) stellar populations.
Indeed, they are well known to obey period-luminosity (PL) and
period-luminosity-color (PLC) relations which are traditionally used to
calibrate secondary distance indicators and, in turn, to estimate the
Hubble constant $H_{0}$ \citep[see e.g.][hereafter DS20 and references therein]{Desomma2020} and \citep[see e.g.][and references therein]{Fiorentino2013,Marconi2005,Riess2011,Riess2019, Ripepi2019}.
Stellar evolution models  predict that CC correspond to
central helium-burning of massive and intermediate-mass stars and obey a
mass-luminosity (ML) relation that is also dependent on chemical
composition, as well as on the efficiency of a number of non canonical physical processes such as rotation, core convective overshooting during the
core hydrogen burning-phase, and mass loss efficiency during (mainly) the red giant branch stage \citep[see, e.g.][and references therein]{chbook, scbook}. The adopted ML relation affects the shape of light curves and radial velocity curves, the coefficients of PLC and period-Wesenheit (PW) relations and, hence, the Cepheid-based distance scale \citepalias[see e.g.][and references therein]{Caputo2005,Desomma2020,Marconi2013,Wood2006}.

By combining the existence of the ML relation for CC with the well
known PL relation and the anti-correlation between mass and age, we can easily conclude that if CC
obey a PL relation, they also have to obey a period-age (PA) relation.
In particular, if the period increases, the
luminosity and the mass also increase according to the PL and the ML
relations, while the Cepheid age decreases.
The existence of a PA relation has been extensively investigated in
literature from both the observational and theoretical point of
view \citep[see e.g.][]{Anderson2016,Efremov1978,EfremovElmegreen1998,Efremov2003, GrebelBrandner1998, Inno2015, Magnier1997,Senchyna2015},
with most of the applications related to cluster pulsators both in the
Milky Way and in other Local Group galaxies, such as M31 and the Magellanic Clouds (MC), for which
independent age estimates were available \citep[see][and references therein]{Bono2005,Efremov2003,Marconi2006}.
However, Cepheid-based ages are more promising than age estimates based, for instance, on isochrone fitting to the cluster color-magnitude diagrams (CMD),
because they only rely on the pulsation period, which can be measured with a high accuracy and is not affected by uncertainties in reddening, distance, and photometric calibration. In addition, the PA relation is also suitable to be applied to field pulsators and, hence, accurate relative\footnote{It is worth noting that relative age estimates are more robust than absolute age determinations, being less affected by systematic uncertainties affecting stellar models.} age estimates, based on this method, can provide strong constraints on the existence of population age gradients in the various Galactic fields\footnote{In this context, we emphasize that the number of CC in the Galactic field is quite larger than in star clusters.}.

From the theoretical point of view, nonlinear convective pulsation
models \citep[][]{Bono2005,Marconi2006} have been adopted to derive
the first completely theoretical PA relations as a function of the
assumed chemical composition.
These authors also showed that, similarly to PL relations, the PA
relation has an intrinsic dispersion that reflects the finite width of the instability strip (IS).
Indeed, it is necessary to include a color term, featuring period-age-color
(PAC) relations, in order to obtain more accurate individual ages
\citep[see e.g.][and references therein]{Bono2005,Marconi2006,Ripepi2017}.
More recently, on the basis of linear nonadiabatic convective Cepheid pulsation models, \citet{Anderson2016} derived theoretical PA relations as
a function of both chemical composition and 
rotation, discussing how the change induced on the ML relation by
variations in the rotation efficiency can significantly modify the
predicted age at a fixed period.
However, as stated above, PAC relations should be preferred to PA
relations to infer individual Cepheid ages and, in turn, to constrain the
star formation history of the Cepheid hosting stellar environments.

In this context, the large sample of  Galactic CC (GCC) for which the \textsl{Gaia} spacecraft provided
the parallaxes and proper motions with unprecedented accuracy, complemented with
multiband photometry, offers a unique dataset to test the predictive capability of theoretical PA and PAC relations, such as the one based on the updated theoretical pulsational framework by \citetalias{Desomma2020}. 
In that paper, we presented a new extended
set of nonlinear convective pulsation models for a solar chemical abundance ($Z$=$0.02$, $Y$=$0.28$) and a wide range of stellar masses and effective temperatures, covering the whole observed
period range of GCC, and varying both the ML relation
and the efficiency of superadiabatic convection.
On this basis, the first predicted light curves, PLC and
PW relations in the \textsl{Gaia} filters were derived and
applied to a selected sample of Gaia Data Release 2 (DR2) GCC \citep[see][for the selection criteria]{Ripepi2019}.

By combining the theoretical predictions for Cepheid periods and colors based on these models with the stellar ages based on the updated stellar evolution predictions in the BaSTI database\footnote{\url{http://basti-iac.oa-abruzzo.inaf.it}} \citep[][]{basti:04,hidalgo:18}, we can derive updated theoretical PA relations and the first PAC relations in the \textsl{Gaia} DR2 bands as a function of both the ML relation and the efficiency of superadiabatic convection. The extension of the pulsational model datebase to a wide range of metallicity and initial helium abundance, mandatory to study extragalactic Cepheids, as well as the implementation of updated prescriptions for the ML relation, will be the subject of an upcoming work (De Somma et al. 2020, in preparation).

Here we focus on the predictions for GCC as age indicators to complement the results presented in \citetalias{Desomma2020} for Cepheids in the Gaia database.
The organization of the paper is as follows. 
In Section 2, we deal with the theoretical scenario \footnote{Hereinafter, with the term “scenario” we refer to a  theoretical framework built by adopting specific assumptions about the physical processes such as mass loss efficiency, core convective overshooting efficiency, etc...} based on recent
evolutionary and pulsation models. In Section 3, we discuss the derivation of updated PA and the first theoretical PAC relations in the \textsl{Gaia} filters, as a function of both the ML and the convective efficiency assumptions. In section 4, the
obtained PAC relations are applied to a sample of GCC in the \textsl{Gaia}
DR2 database to obtain individual ages and, in turn, the predicted age
distribution. The conclusions and some future developments close the paper.

\section{A theoretical scenario for Galactic Cepheids}

\subsection{The evolutionary framework}

Our previous works on the theoretical PA relationship \citep[see,][]{Bono2005,Marconi2006} were based on the theoretical evolutionary framework developed by \cite{basti:04}, named the BaSTI library. However, since the first release of the BaSTI library, several improvements to the stellar physics inputs, some revisions of the solar metal distribution, and corresponding revisions of the solar metallicity \citep[we refer to, e.g.][and references therein for a detailed discussion on this issue]{bs:14} have become available. Therefore, the BaSTI library has been recently updated. This version is known as the BaSTI-IAC version\footnote{The whole BaSTI-IAC library is available at the following URL: http://basti-iac.oa-abruzzo.inaf.it}, and takes into account all the presently available updates and improvements in the input physics. The new complete library for a scaled solar chemical mixture has been provided in \cite{hidalgo:18}, and we refer the interested reader to this reference for a detailed discussion on the physical framework adopted for the model computations. This notwithstanding, for the sake of completeness, it is important to briefly discuss here the main differences between the BaSTI-IAC models adopted in the present work and BaSTI predictions accounted for in \cite{Bono2005} and \cite{Marconi2006}.

One of the main differences is associated with the use of different reference solar heavy element distributions. The previous BaSTI models were based on the solar mixture provided by \cite{gs:98}. The new ones take into account the recent, significant revisions of the solar metal distributions, and rely on measurements provided by \cite{caffau:11}, supplemented when necessary  by the abundance estimates provided by \cite{lodders}.

The treatment of overshooting beyond the Schwarzschild boundary of convective cores during the H-burning stage, is included as an instantaneous mixing between the formal convective border and layers at a distance $\lambda_{ov}H_P$ from this boundary – keeping the radiative temperature gradient in this overshooting region. $H_P$ is the pressure scale height at the Schwarzschild boundary, and $\lambda_{ov}$ is a free parameter that in the BaSTI-IAC model computations has been set equal to 0.2, decreasing to zero when the mass decreases below a certain value\footnote{We refer to \cite{hidalgo:18} for a discussion about the need to decrease the core convective overshooting with the mass as well as the procedure adopted for reducing the value of $\lambda_{ov}$ below a given critical mass.} (typically, equal to $\sim1.5M_\odot$). As the previous BaSTI library, the new BaSTI-IAC library provides two complete sets of evolutionary models alternatively accounting (non canonical models) or not accounting at all for core convective overshooting (canonical scenario).



The mass loss phenomenon is accounted for by using the \cite{reimers} formula, with the free parameter $\eta$ - present in this mass loss prescription - set equal to zero when neglecting mass loss, or to $\eta$=0.3 when mass loss is accounted for \citep[we refer to][for the discussion about this choice]{hidalgo:18}.

The treatment of superadiabatic convection is based on the \cite{bv:58} flavor of the mixing length theory, using the formalism by \cite{cg:68}. 
The value of the mixing length free parameter, fixed via the standard solar model (SSM) calibration\footnote{The adopted procedure and the properties of the BaSTI-IAC SSM are fully described in \cite{hidalgo:18}.}, to $\alpha_{ml}$  = 2.006, was kept constant for all stellar masses, initial chemical compositions and evolutionary phases.

As mentioned, the calibration of the SSM sets the value of $\alpha_{ml}$, and the initial solar He abundance $Y_\odot$ and metallicity $Z_\odot$: the BaSTI-IAC SSM -- properly accounting for atomic diffusion of both He and metals -- matches the empirical solar constraints with the following initial abundances $Z_\odot$=$0.01721$ and $Y_\odot$=$0.2695$.

For the present analysis, we selected the stellar models with the solar chemical composition and a mass range between $4M_\odot$ and $11M_\odot$, with a step of $0.5M_\odot$.

\begin{figure*}
\includegraphics[width=\textwidth]{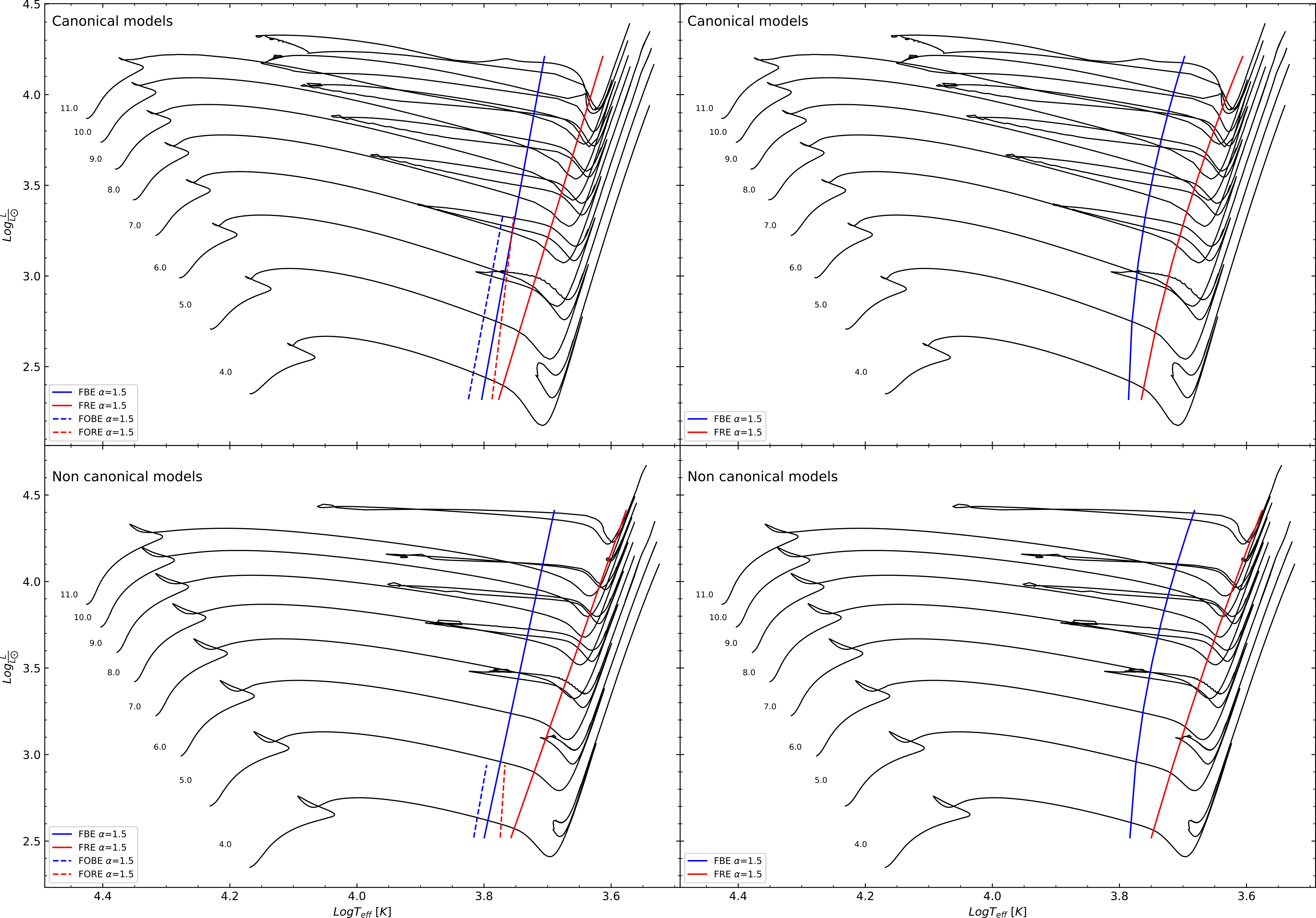}
\caption{The location in the HR diagram of the selected massive and intermediate-mass stellar models for the adopted solar chemical composition, compared with the predicted linear IS boundaries ({\sl left panels}) and quadratic IS boundaries ({\sl right panels}) of radial F (solid line) and FO-mode pulsators (dashed line) as obtained for the canonical  {\sl (top panels)} and non canonical ML relation {\sl (bottom panels)}, assuming $\alpha_{ml}~$ = $1.5$. (see text for more details).}
\label{fig:hr_diagram}
\end{figure*}

\subsection{The pulsational framework}

The pulsation models used in this work are included in the updated
dataset of nonlinear and convective pulsation models computed for
the chemical composition\footnote{We note that the small difference both in the initial metallicity and helium abundance between the evolutionary stellar model set and the pulsational one is expected to have a negligible, if any, effect on the predicted pulsational properties. However, in order to check this issue we computed a few, additional pulsational models by adopting the exact chemical pattern used for the evolutionary computations. As a result we verified that the small differences in the adopted values for Z and Y do not affect the theoretical pulsational framework.} $Z$ = $0.02$, $Y$ = $0.28$, by
\citetalias{Desomma2020}. By adopting the automated procedure described in that paper, we produced an extended dataset of fundamental (F) and first overtone (FO) models covering a wide range of input parameters. 

For the first time a fine grid of pulsation models by simultaneously varying both the ML relation and the efficiency of superadiabatic convection was computed.
More in detail, three ML relations were adopted: the canonical ML relation derived in \citet{BonoTornambe2000} (named {\sl case A}), obtained by
neglecting mass loss, rotation and overshooting, a non canonical ML relation obtained by increasing the canonical luminosity level by $\Delta\log(L$/$L_\odot)$=$~0.2$ (hereinafter {\sl case B})  and, a still brighter ML relation obtained by increasing the canonical luminosity level by $\Delta\log(L$/$L_\odot)$=$~0.4$ (named {\sl case C}).
The cases B and C for the ML relationship take into account the increase of the luminosity corresponding to the blue loop Cepheid stage, at a given stellar mass, induced by an increase of the mass size of the convective core during the central H-burning stage as due to non canonical processes such as convective overshooting and/or rotational induced mixing \citep[we refer to][for more details on this issue]{Bono2005}.

We note that, despite the improvements in the physical framework adopted for the BaSTI-IAC stellar model computation, the ML relationship predicted by the solar chemical composition, canonical (no overshooting) models is in quite good agreement with the the canonical ML relation adopted to build the Cepheid pulsation models \citep{BonoTornambe2000}. At the same time, the increase of the brightness of the ML relationship by $\Delta\log(L$/$L_\odot)$=$~0.2$ at a given stellar mass to simulate the effect of non canonical processes is consistent with the brightness increase of the ML relation predicted by the non canonical, BaSTI-IAC models with respect the corresponding canonical ones \citep[we address the interested reader to][for a detailed comparison between the various model libraries]{hidalgo:18}. These considerations allow us to safely rely on the ML prescriptions based for developing the pulsational scenario adopted in the present work.

As for the superadiabatic convective efficiency, three values of the
mixing length parameter\footnote{We note that there are some differences in the implementation of the mixing length theory in an evolutionary code \citep[see, e.g.][]{cassisiml} and pulsational code adopted in the present analysis \citep[see, e.g., the discussion presented in][]{bs:94}. In the latter the mixing lenght parameter ($\alpha_{ml}$=$~l$/$H_{P}$ where l is the length of the path covered by the convective elements and $H_{P}$ is the local pressure scale height), is adopted to close the nonlinear system of dynamical and convective equations \cite[see][for details]{bs:94}. As a consequence, the value of the free parameter $\alpha_{ml}$ adopted in the evolutionary computations can not be directly compared to the ones adopted in the pulsational modelling.} were chosen; namely, $\alpha_{ml}$~=$1.5$ $\alpha_{ml}~$=$1.7$ and $\alpha_{ml}~$=$1.9$. 

For the purpose of the current work, we considered only models with ML
relations corresponding to {\sl cases A} and {\sl B}, and two assumptions on the efficiency of convection in the outer layers, namely $\alpha_{ml}~$=$1.5$ and
$\alpha_{ml}~$=$1.7$. The reasons for not considering the brightest models and the highest $\alpha_{ml}$ value are related to the
need for consistency with the adopted evolutionary scenario that
do not predict overluminosities as high as +0.4 dex with respect to the canonical ML, and the small number of pulsating models obtained by \citetalias{Desomma2020} for $\alpha_{ml}~$=$1.9$ \citepalias[see][for details]{Desomma2020}. 

The input parameters adopted for F and FO-mode models are listed in Tables 1 and 2 in \citetalias{Desomma2020}, respectively. \footnote{A machine-readable version of the aforementioned Tables is available through the following link: \url{https://iopscience.iop.org/article/10.3847/1538-4365/ab7204}.}
For each assumed combination of mass, luminosity and $\alpha_{ml}$, the F and FO
IS boundaries were derived by
\citetalias{Desomma2020}. The linear fit of the inferred F and FO boundaries as a function of the luminosity level is reported in Tables 5 and 6 in \citetalias{Desomma2020}, while the quadratic fit coefficients are listed in current Table \ref{tl_f_quad}. In the case of the FO-mode, only the linear fit was taken into account due to the smaller number of pulsating models and the well known linear behaviour of the predicted edges \citep[see e.g.][and references therein]{Bcm2000}.

Figure \ref{fig:hr_diagram} shows the location in the HR diagram of the linear (left panels) and quadratic (right panels) fits of the IS boundaries with respect the
evolutionary tracks discussed in subsection 2.1. For clarity reason,
only a subsample of stellar evolutionary tracks was plotted. Inspection of this figure suggests that stellar models with mass lower than about 5$M_{\odot}$, cross the predicted IS only once (first crossing), while evolving towards the Red Giant Branch. However, stars with masses higher than $\sim5M_{\odot}$ show three crossings with the second and the third crossings corresponding to the blue-ward and red-ward evolution along the blue loop, respectively.

As expected on the basis of evolutionary considerations, the time spent during the first crossing is significantly shorter than the second and the third one. For example, in the case of a 6$M_{\odot}$ star the time spent inside the instability strip varies from $7.41\cdot10^3 yr$ for the ${\mathrm 1^{th}}$ crossing to $3.73\cdot10^5 yr$ for the ${\mathrm 2^{nd}}$ and $1.80\cdot10^5 yr$ for the ${\mathrm 3^{rd}}$ one; while for a 11$M_{\odot}$ model the corresponding evolutionary lifetimes are $2.87\cdot10^3$ $yr$ for the ${\mathrm 1^{th}}$ crossing, and $3.16\cdot10^4 yr$ and $1.41\cdot10^4 yr$ for the ${\mathrm 2^{nd}}$ and ${\mathrm 3^{rd}}$ crossing, respectively.

For the reliability of evolutionary predictions about the blue loop
morphology (and extension), it is worth mentioning that the physical reasons for the blue loops challenged for long time our understanding of stellar evolution, and it is still a difficult task to predict the response of an intermediate-mass stellar model, during this phase, to changes in the physical parameters and/or the physical assumptions adopted in the
evolutionary computations. Indeed, both the morphology, and the actual occurrence of the blue loops have a highly non-linear dependence on the physical inputs and assumptions adopted in the stellar evolution models. Even minor changes, for example, in the chemical composition, initial mass, mixing process - such as convective core and envelope overshooting - efficiency can have huge impact on the blue loop properties \citep[we refer to, e.g.,][and references therein for a detailed discussion on this topic]{renzini:92,scbook}.

\begin{table*}
\caption{\label{tl_f_quad} The coefficients of the quadratic relation $\log T_{eff}$ = a + b \lsun +c (\lsun)$^2$ for the boundaries of the F-mode IS, varying both the ML relation and the mixing length parameter.}
\centering
\begin{tabular}{ccccccccc}
\hline\hline
$\alpha_{ml}$&ML&a&b&c&$\sigma_{a}$&$\sigma_{b}$&$\sigma_{c}$&$R^2$\\
\hline
FBE\\
\hline
1.5&A&3.667&0.104&-0.023&0.054&0.033&0.005&0.981\\
1.5&B&3.675&0.098&-0.022&0.053&0.031&0.004&0.988\\
1.7&A&3.708&0.092&-0.023&0.037&0.023&0.003&0.994\\
1.7&B&3.820&0.019&-0.012&0.069&0.040&0.005&0.986\\
\hline
FRE\\
\hline
1.5&A&3.788&0.030&-0.018&0.031&0.019&0.003&0.998\\
1.5&B&3.901&-0.042&-0.007&0.150&0.087&0.012&0.968\\
1.7&A&3.773&0.044&-0.018&0.050&0.031&0.005&0.993\\
1.7&B&3.645&0.121&-0.030&0.042&0.025&0.003&0.997\\
\hline\hline
\end{tabular}
\end{table*}

\section{Tools for deriving Galactic Classical Cepheids ages}

\subsection{The theoretical Period-Age relation}
In order to derive theoretical PA relations, we combined the evolutionary tracks with the predicted instability strips and pulsation periods inferred from the pulsation models computed in \citetalias{Desomma2020}.

At first, we adopted two different sub-samples: one including
all the crossings and the other including only the second and third
crossings. However, since we verified that the resulting PA relations are not
significantly affected by the choice of specific sub-sample, in the following we only consider the relations obtained including all the crossings\footnote{As discussed at the end of the previous section, this occurrence is due to the fact that the time spent in the IS during the first crossing is a small fraction of the time spent inside the IS during the blue loop evolutionary stage.}. Incidentally, we note that this choice is also consistent with the methodological approach adopted by \citet{Bono2005} and \citet{Marconi2006} to derive the PA relationship.

Adopting the period-luminosity-mass-temperature (PMLT) relation  in the form $\log P$ = a + b$\log T_{eff}$ + c $\log$ \msun + d \lsun, derived by \citetalias{Desomma2020}, we were able to predict the period for each 
combination of mass, luminosity and effective temperature along the
selected portions of the evolutionary tracks. By combining the period estimate with the age predicted by the evolutionary models, we directly
derived the theoretical PA relations through a linear regression procedure.
The coefficients obtained for each combination of ML and
convective efficiency, assuming both linear and quadratic fits to the F
boundaries, are reported in Tables \ref{pa_f_fo_lin} and \ref{pa_f_quad}, respectively. We notice that, due to the limited number of FO-mode models, in this case only the PA relations based on linear boundaries were obtained.

\begin{figure*}
\includegraphics[width=\linewidth]{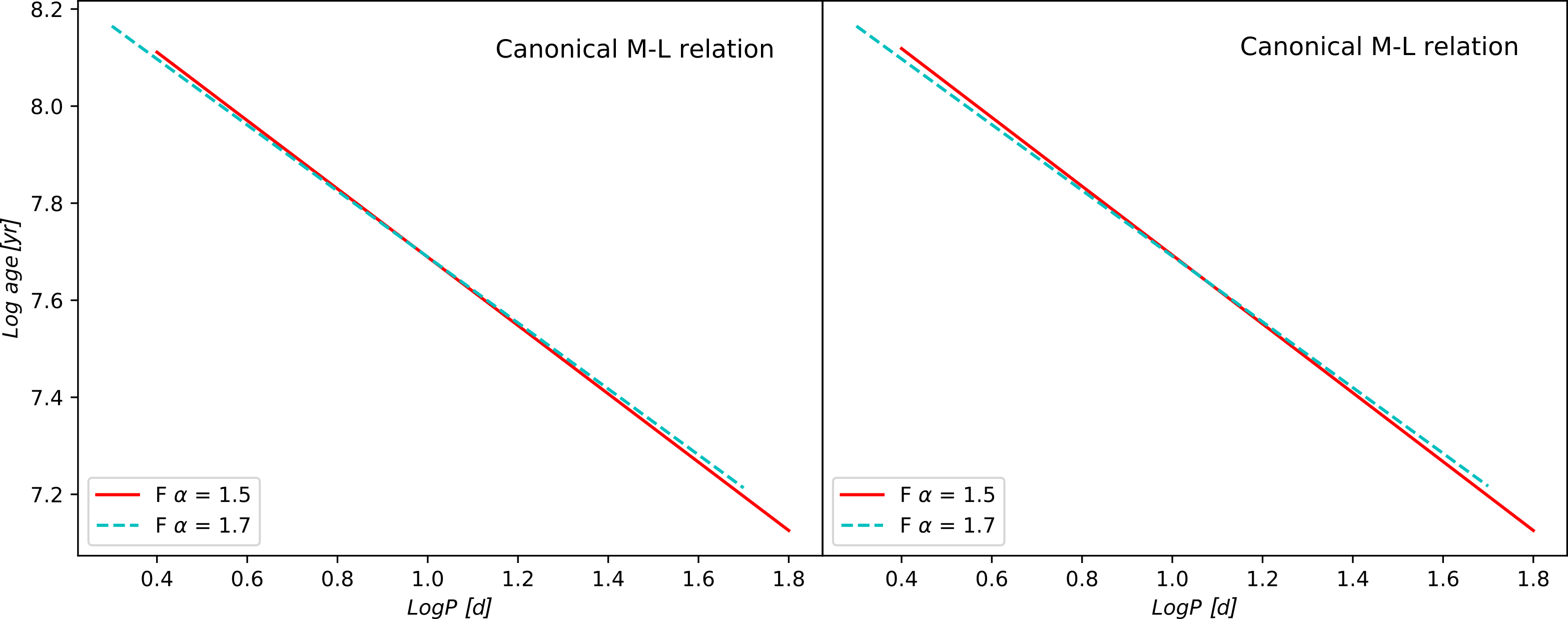}\par
\caption{\label{fig:pa_lin_alfa_comp} Canonical PA relations for the various assumptions about superadiabatic convective efficiency, assuming linear {\sl (left panel) and quadratic {\sl (right panel)}} analytical relations for the IS boundaries.}
\end{figure*}

Figure~\ref{fig:pa_lin_alfa_comp} shows a comparison between the predicted PA relations for F pulsators with $\alpha_{ml}~$=$1.5$ and $\alpha_{ml}~$=$1.7$, in the {\sl case A} ML relation and assuming both the quadratic (right panel) and the linear (left panel) boundary analytical relations.

The data shown in this figure reveal that the PA relation is largely unaffected by the exact value of the mixing length parameter adopted in the pulsational model computations; therefore, one can safely assume that the predicted Cepheid ages are barely affected by the lingering uncertainties in the adopted mixing length parameter.

\begin{figure*}
\includegraphics[width=\linewidth]{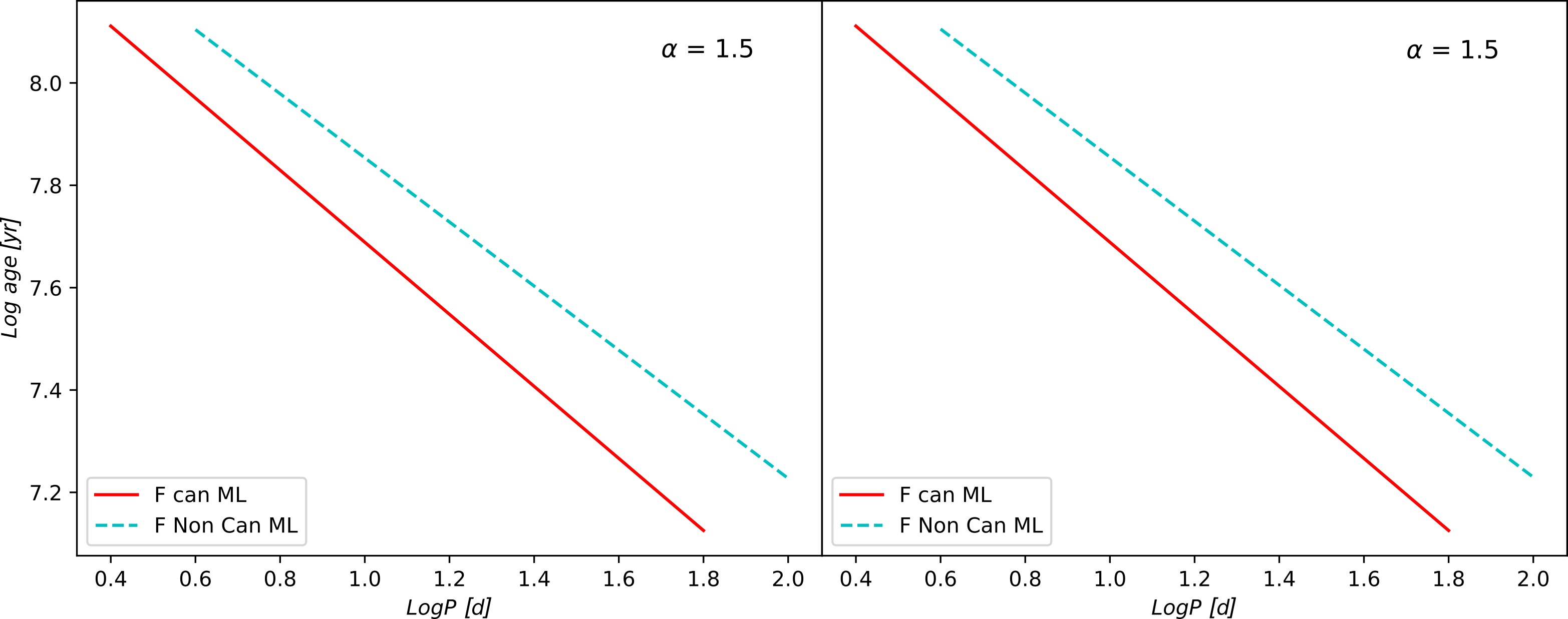}\par
\caption{\label{fig:pa_lin_ml_comp} Fundamental PA relations at a fixed
mixing length parameter $\alpha_{ml}~$=$1.5$ for the two assumed ML relations, assuming linear {\sl (left panel)} and quadratic {\sl (right panel)} IS boundary analytical relations.}
\end{figure*}

Figure~\ref{fig:pa_lin_ml_comp} shows the same kind of
comparison but varying the ML relation at a fixed $\alpha_{ml}~$=$1.5$.
In this case, we notice that brighter ML relations provide systematically older ages. This result is consistent with the theoretical prediction of a longer core hydrogen burning stage for stellar models accounting for core convective overshooting as the ones that have been combined with our {\sl case B} pulsation models; an occurrence obviously due to the larger amount of fuel available during the central H-burning stage.
In particular, the age difference between {\sl case B} and {\sl A}
ranges from $\sim$ 36 \% at $\log{P}$=$0.4$ to $\sim$ 60 \%  at $\log{P}$=$1.8$. We notice that the period range in these figures is
the same as that estimated by \citetalias{Desomma2020}
(we refer to the quoted reference for more details) for {\sl cases A} and {\sl B} models. As the relations obtained assuming linear and quadratic F boundaries
are found to be perfectly consistent within the errors, in the following we only rely on the relations obtained assuming linear boundaries.

\begin{table*}
\caption{\label{pa_f_fo_lin}The coefficients of the F
and FO PA relations in the form $\log t$ = $a$ + $b \log P$, assuming linear IS boundaries and adopting  both {\sl case A} and {\sl B} ML relations and $\alpha_{ml}~$=$1.5$ and $\alpha_{ml}~$=$1.7$. The last two columns represent the R-squared ($R^2$) and the root-mean-square deviation ($\sigma$) coefficients.}
\centering
\begin{tabular}{cccccccc}
\hline\hline
Fundamental mode\\
\hline
$\alpha_{ml}$&ML&a&b&$\sigma_{a}$&$\sigma_{b}$&$R^2$&$\sigma$\\
\hline
1.5&A&8.393&-0.704&0.008&0.009&0.916&0.084\\ 
1.7&A&8.369&-0.680&0.015&0.017&0.908&0.080\\
1.5&B&8.480&-0.626&0.010&0.009&0.866&0.080\\
1.7&B&8.460&-0.618&0.013&0.010&0.852&0.090\\
\hline
First overtone mode\\
\hline
1.5&A&8.120&-0.396&0.020&0.057&0.506&0.052\\
\hline\hline
\end{tabular}
\end{table*}

\begin{table*}
\caption{\label{pa_f_quad}The coefficients of the F-mode PA
relation $\log t$ = $a$ +$b \log P$, assuming quadratic IS
boundaries and adopting both {\sl case A} and {\sl B} ML relations and $\alpha_{ml}~$=$1.5$ and $\alpha_{ml}~$=$1.7$. The last two columns represent the R-squared ($R^2$) and the root-mean-square deviation ($\sigma$) coefficients.}
\centering
\begin{tabular}{cccccccc}
\hline\hline
Fundamental mode\\
\hline
$\alpha_{ml}$&ML&a&b&$\sigma_{a}$&$\sigma_{b}$&$R^2$&$\sigma$\\
\hline
1.5&A&8.396&-0.708&0.007&0.009&0.910&0.083\\
1.7&A&8.362&-0.676&0.014&0.017&0.898&0.084\\
1.5&B&8.511&-0.660&0.010&0.009&0.856&0.090\\
1.7&B&8.477&-0.639&0.013&0.010&0.847&0.099\\
\hline\hline
\end{tabular}
\end{table*}

\begin{figure*}
\begin{multicols}{2}
\includegraphics[width=\linewidth]{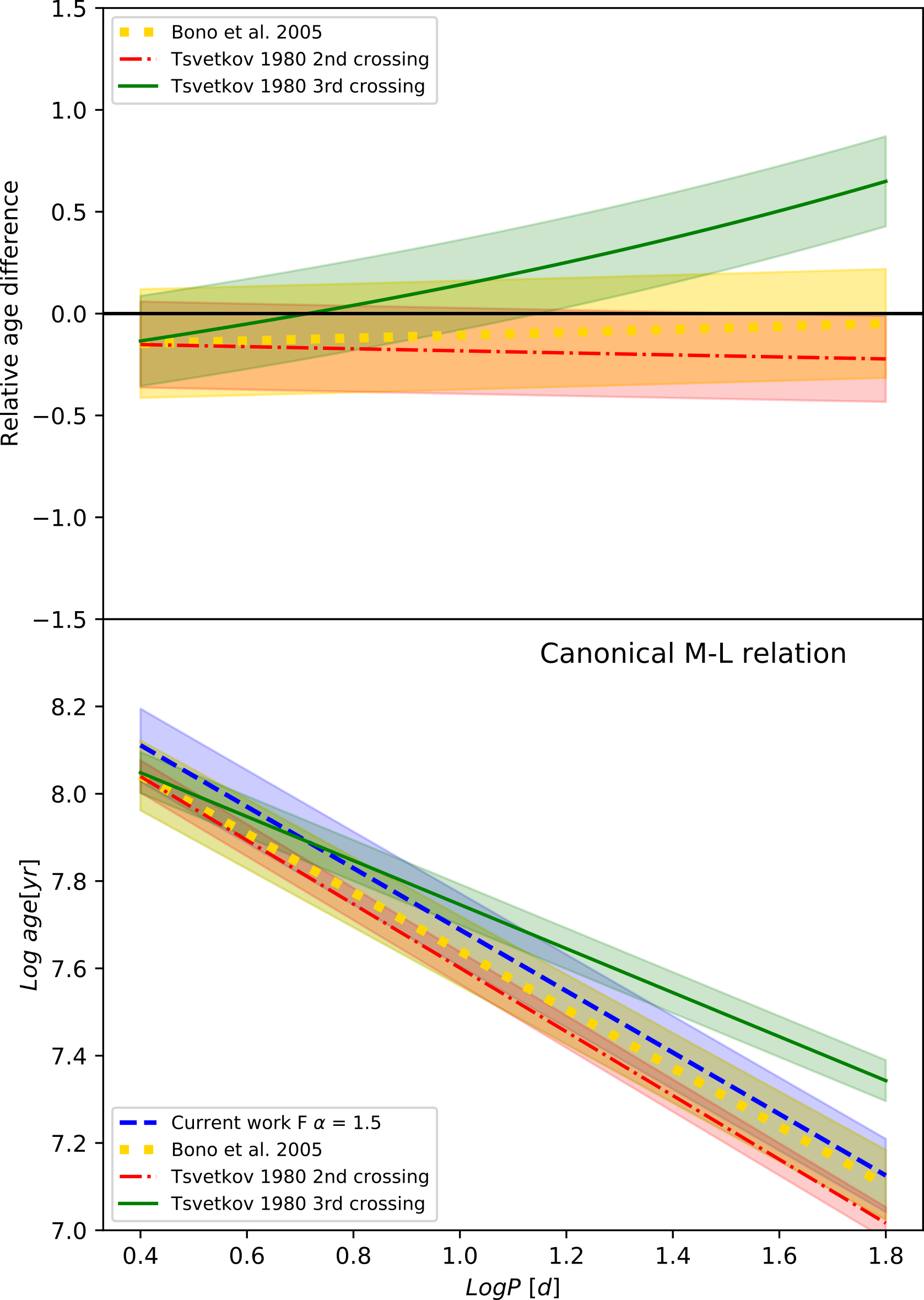}\par 
\includegraphics[width=\linewidth]{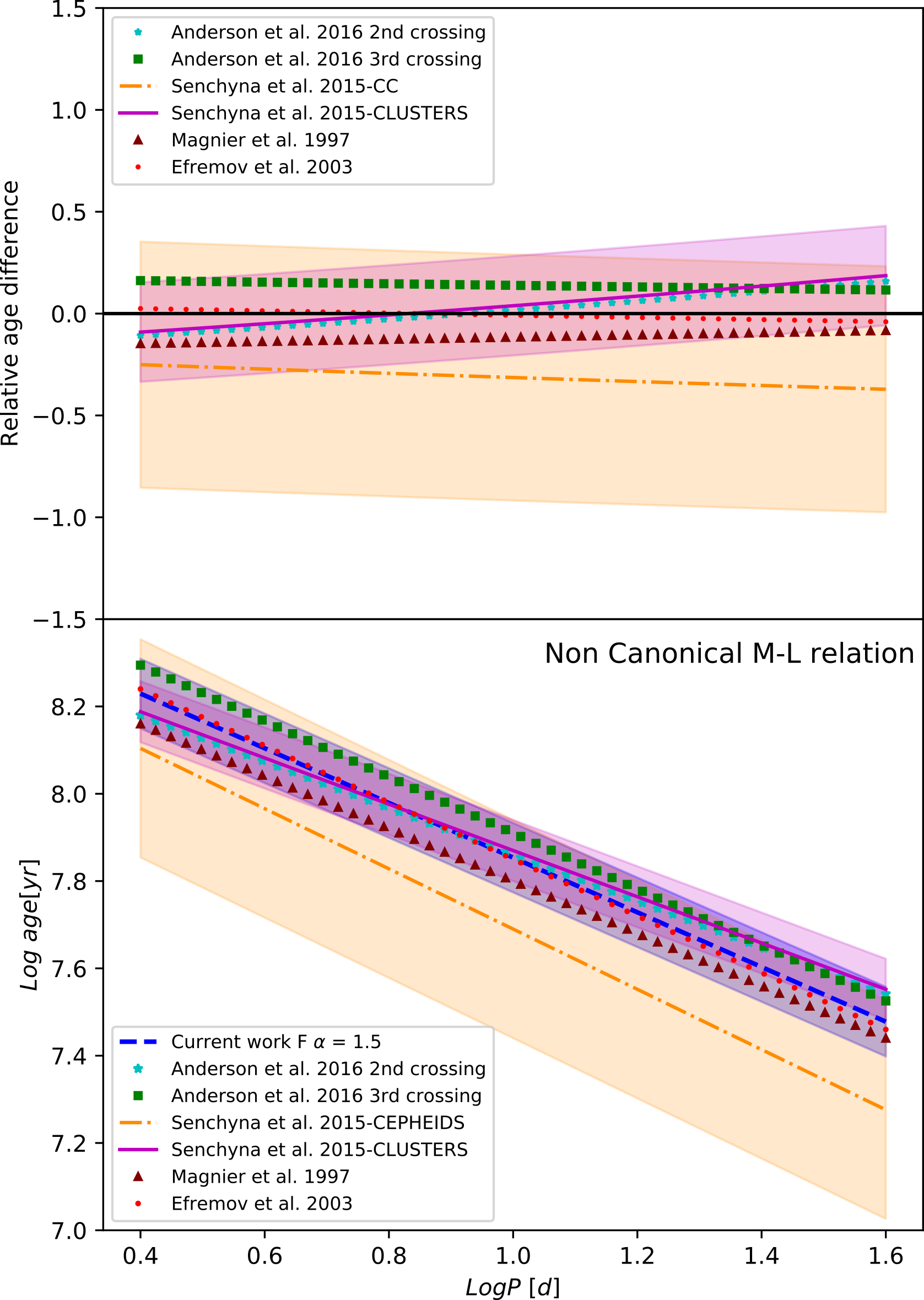}\par
\end{multicols}
\caption{\label{fig:pa_comparison} {\sl Bottom panels}: comparison between present F-mode PA
relations (dashed blue line), as obtained by varying the adopted ML relation (see labels) but at a fixed value for the mixing length ($\alpha_{ml}$), with similar predictions from the literature: the dashed red line and dash-dotted green line show the PA relations by \citet{Tsvetkov1980}, for the second and third crossings, respectively; the dotted yellow line shows the PA relation obtained by \citet{Bono2005}; the relations marked with stars and squares represent the PA relations by \citet{Anderson2016}, for the second and third crossings, respectively; the dash-dotted orange line refers to the PA relation by \citet{Senchyna2015} for M31 CC ; the solid magenta line is the PA relation by \citet{Senchyna2015} for M31 clusters; filled circles and triangles correspond to the PA relations by \citet{Efremov2003} and \citet{Magnier1997}, respectively. The colored shaded areas represent the $1\sigma$ errors on these relationships as provided by the authors. {\sl Upper panels}: the relative age difference between the age predictions obtained by present PA relations for the canonical case {\sl (left panel)} and the non canonical one {\sl (right panel)} and those obtained with the mentioned PA relationships taken from the literature.}
\end{figure*}

\subsubsection{Comparison with the literature}

In Figure~\ref{fig:pa_comparison} we compare our theoretical PA
relations (dashed blue line) obtained for {\sl case A} (left panels) and {\sl B} ML (right panels)- by assuming in both cases $\alpha_{ml}~$=$1.5$, with other theoretical relations available in the
literature. We selected a common period range for both {\sl case A} (from
$\log P~$=$~0.4$ to $\log P~$=$~1.8$) and {\sl B} (from $\log
P~$=$~0.4$ to $\log P~$=$~1.6$). 

For the canonical case (left panels), we
compare our {\sl case A} relations with: i) the theoretical PA relation published by \citet{Bono2005} (dotted yellow line) and ii) the semi-empirical PA relations derived by \citet{Tsvetkov1980} for the second (dashed red line) and third crossings (dash-dotted green line), respectively. For the non canonical case (right panels), we compare our {\sl case B} relations with: i) the
theoretical PA relations derived by \citet{Anderson2016} assuming a mild rotation efficiency, for the second (cyan star marker) and third crossings (green square marker), respectively; ii) the semi-empirical PA relations by \citet{Senchyna2015} based both on M31 Cepheids (dash-dotted orange line) and on M31 cluster isochrone fitting (solid magenta line); iii) the semi-empirical relation provided by \citet{Magnier1997} (brown triangle marker) and iv) the empirical relationship by \citet{Efremov2003} (red dot marker).

\begin{figure*}
\includegraphics[width=\columnwidth]{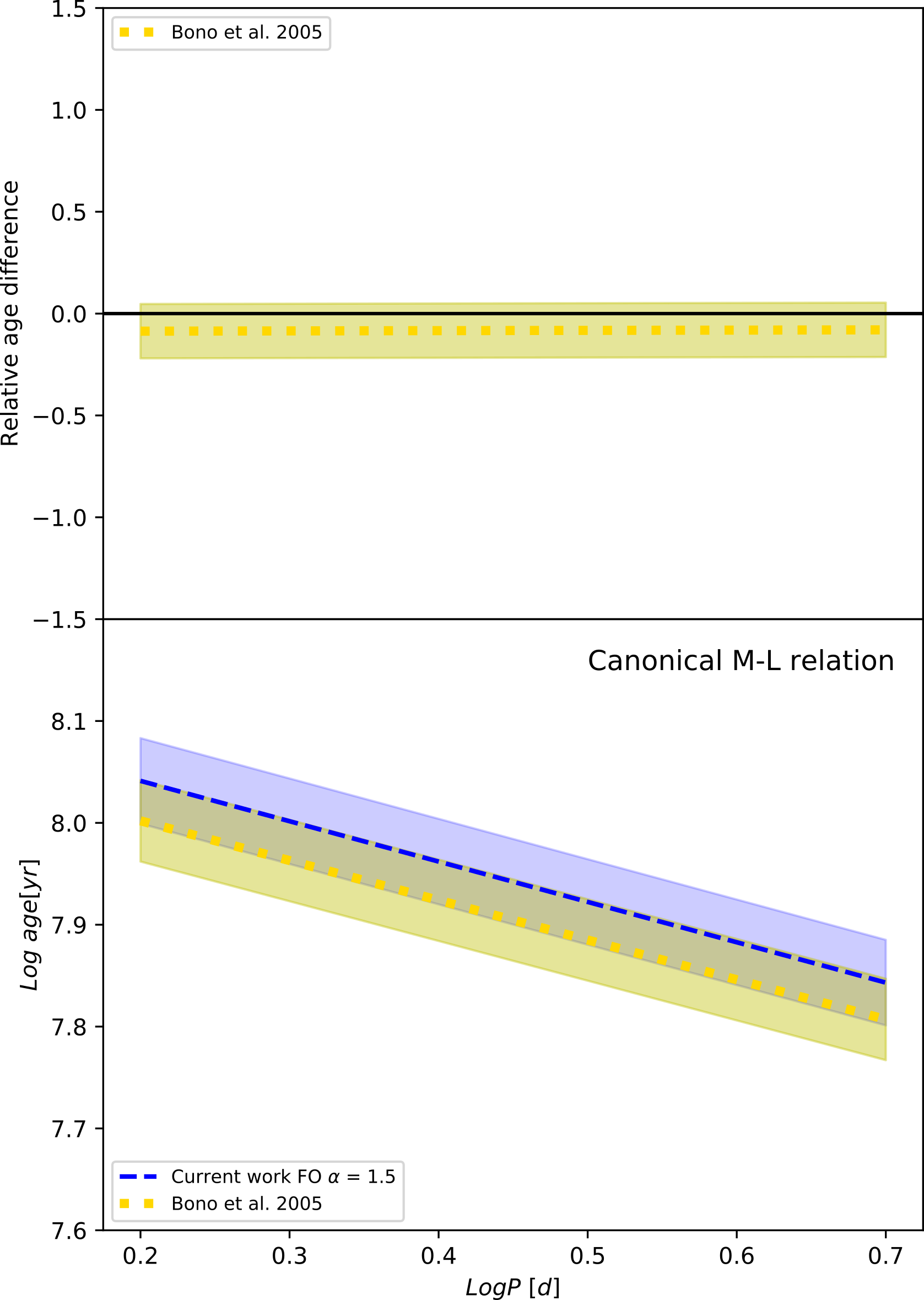}\par
\caption{{\sl Bottom panel}: comparison between present FO-mode PA relation (dashed blue line) obtained for ML case A and $\alpha_{ml}~$=$1.5$, with the theoretical FO-mode GCC PA relation obtained by \citet{Bono2005} (dotted yellow line). {\sl Upper panel}: the relative difference between the age estimates provided by these PA relations.}
\label{fig:pa_FO_comparison}
\end{figure*}

Figure~\ref{fig:pa_FO_comparison} shows the same kind of comparison but for {\sl case A} FO-mode models with $\alpha_{ml}~$=$ 1.5$ and the FO relation by \citet{Bono2005}. 

The coefficients of the various PA relations used for the comparison are summarized in Table~\ref{pa_source}.

Inspection of the bottom panels of Figures~\ref{fig:pa_comparison} and \ref{fig:pa_FO_comparison} suggests a very good agreement, within the errors, between our canonical PA relations and the one previously derived by \citet{Bono2005} on the basis of a less extended and updated set of models and a
slightly different ML relation. No error estimate was provided by \citet{Anderson2016}, \citet{Efremov2003} and \citet{Magnier1997} but we found a good agreement between their (all non canonical) relations and our case B.

In order to better quantify the level of agreement among the various PA relationships, the two upper panels in Figure \ref{fig:pa_comparison} show the relative age difference between the relations obtained in the present work and the ones selected from the literature for both the canonical and non canonical ML cases. For the canonical case, present PA relation predicts ages systematically larger than the other relations with a maximum difference of the order of $\sim15-20$\%. The smallest difference -- well within the associated intrinsic dispersions --
is found when considering the PA relation provided by \citet{Bono2005}, while the maximum discrepancy is obtained when comparing the present result with the relation obtained
by \citet{Tsvetkov1980} for the third crossing. In this case the difference can also reach a value as large as $\sim50$\% and it can be likely related to the difference in luminosity between the second
and the third crossings in spite of the quite similar evolutionary times.

For the non canonical case, the relative age differences are within $\sim15-20$\%, apart from the case of the PA relation, based on Cepheids in M31, provided by \citet{Senchyna2015}, that predicts ages about 25\% smaller than those provided by our relation. However, we wish to emphasize the large uncertainty associated to the \citet{Senchyna2015} Cepheid-based PA relationship. For the case of FO PA, the agreement is remarkably good: there is a difference of about 5\% between present predictions and those obtained by using the \citet{Bono2005} relation, but again well within the errors.

The general good agreement among the various PA relations based on similar ML relations, as well as the age difference obtained when comparing the results based on canonical and non canonical stellar models support the idea that different assumptions on non canonical physical processes, e.g., moderate rotation as in \citet{Anderson2016}, and
mild  overshooting as in the other semi-empirical relations, have the same effect of producing a brighter luminosity at a fixed mass, and in turn, an older age at a fixed period. 

\begin{table}
\caption{\label{pa_source}The PA relations $\log t$ = $a$ +$b \log P$ derived by various authors for different CC samples. $\sigma$ is the predicted root-mean-square deviation coefficient.}
\centering
\begin{tabular}{ccccc}
\hline\hline
authors & source & a & b & $\sigma$\\
\hline
\citet{Anderson2016}&F GCC 2nd crossing&8.393&- 0.532&\\
\citet{Anderson2016}&F GCC 3rd crossing&8.551&-0.641\\ 
\citet{Bono2005}&F GCC&8.31&-0.67&0.08\\ 
\citet{Bono2005}&FO GCC&8.08&-0.39&0.04\\ 
\citet{Efremov2003}&LMC bar CC&8.50&-0.65&\\
\citet{Magnier1997}&M31 CC&8.4&-0.6&\\
\citet{Senchyna2015}&M31 CLUSTERS&8.40&-0.53&0.07\\
\citet{Senchyna2015}&M31 CC&8.38&-0.69&0.25\\
\citet{Tsvetkov1980}& F GCC 2nd crossing&8.332&-0.731&0.037\\
\citet{Tsvetkov1980}& F GCC 3rd crossing&8.250&-0.504&0.047\\ \hline\hline
\end{tabular}
\end{table}

\subsection{The theoretical Period-Age-Color relation in the \textsl{Gaia} filters}

PA relations have the advantage to allow a direct
evaluation of CC ages when only the pulsation periods are known, however they suffer from the limitation related to the finite color width of the IS, that causes - mostly in the optical photometric passbands - an inherent dispersion of PL relations. This occurrence implies that PA relations are also affected by an intrinsic scatter with a range of colors and periods for each fixed age; an effect than can be removed if a color term is included in the linear regression procedure.

As the main objective of the present work was to derive the individual
ages of GCC with Gaia DR2 parallaxes, we used the Gaia
band model light curves derived in \citetalias{Desomma2020} and the
resulting mean magnitudes $<G_{BP}>$ and $<G_{RP}>$, to compute the
first PAC relations in the \textsl{Gaia} filters. Table \ref{pac_f_fo}
lists the coefficients of the derived PAC relations, for the F and FO-mode models and the discussed assumptions concerning ML and $\alpha_{ml}$. 

\begin{figure}
\includegraphics[width=\linewidth]{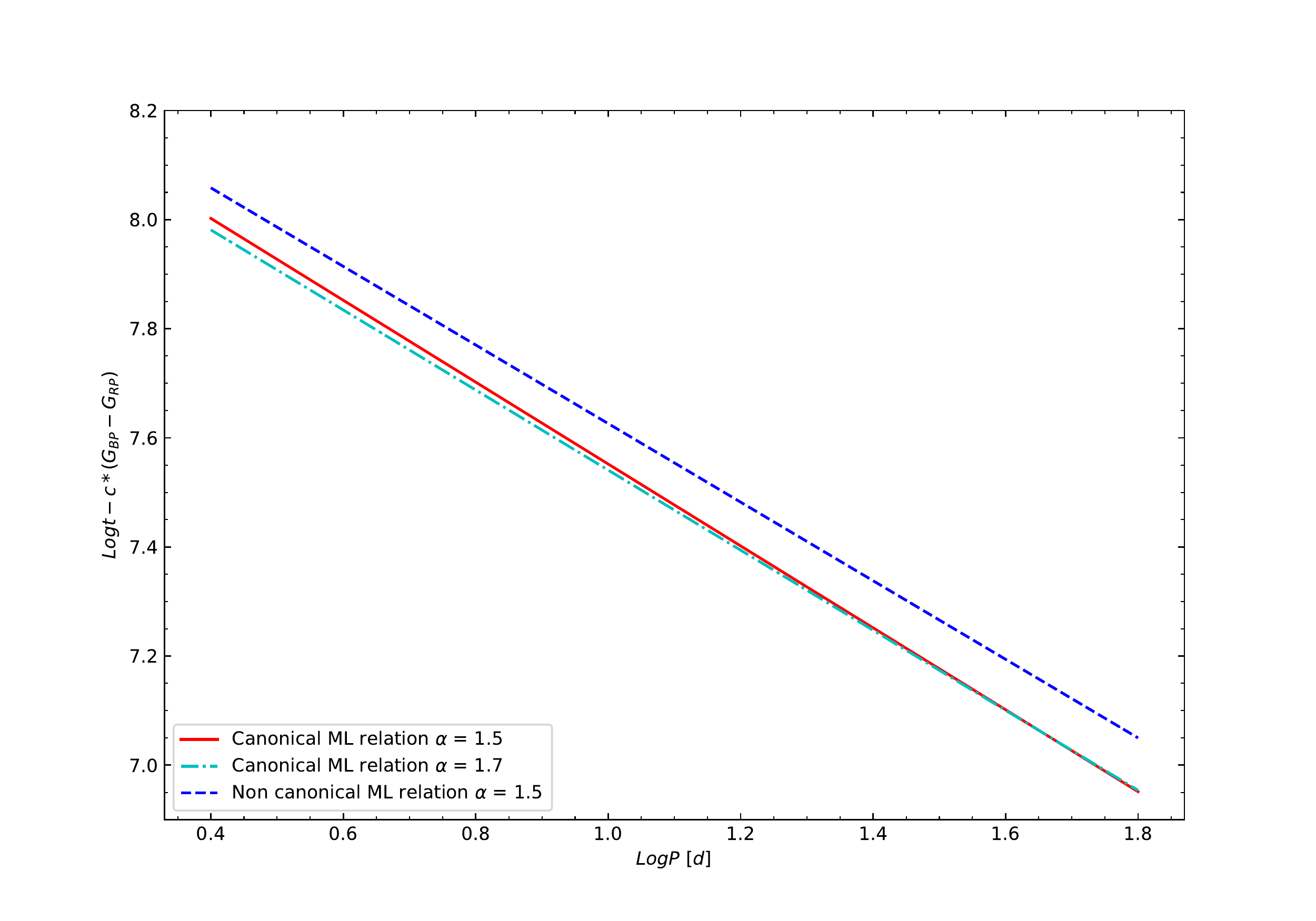}\par
\caption{Projection onto a plane of the new Gaia band PAC relation (red solid line) for F-mode models as obtained by adopting case A ML and $\alpha_{ml}~$=$1.5$; for comparison the same relation as obtained by varying the super adiabatic convective efficiency (dash-dotted cyan line) or the ML relation (dashed blue line) is also shown.}
\label{fig:pac_alfa_comp}
\end{figure}

In Figure \ref{fig:pac_alfa_comp}, we compare the projections onto a plane of the PAC relations varying the assumptions on $\alpha_{ml}$ and the ML relations. As already found for the PA relations, a change in the super adiabatic convection efficiency does not affect the ages predicted from the new derived theoretical PAC relations. As a consequence, in the following we only adopt the PAC relation obtained for $\alpha_{ml}~$=$1.5$.

\begin{table*}
\caption{\label{pac_f_fo} The coefficients of the PAC relation: $\log t$ = $a$ + $b \log P$ + $c$ ($<G_{BP}>$ - $<G_{RP}>$), for F and FO pulsators, by varying both the ML relation and mixing length value.}
\centering
\begin{tabular}{cccccccccc}
\hline\hline
Fundamental mode\\
\hline
$\alpha_{ml}$&ML&a&b&c&$\sigma_{a}$&$\sigma_{b}$&$\sigma_{c}$&$R^2$&$\sigma$\\
1.5&A&8.303&-0.751&0.121&0.045&0.025&0.060&0.916&0.083\\
1.7&A&8.346&-0.720&0.077&0.109&0.062&0.156&0.934&0.046\\
1.5&B&8.275&-0.734&0.278&0.029&0.017&0.037&0.876&0.077\\
1.7&B&8.453&-0.624&0.012&0.030&0.026&0.047&0.852&0.090\\
\hline
First overtone mode\\
\hline
1.5&A&7.961&-0.508&0.255&0.137&0.064&0.204&0.603&0.046\\
\hline\hline
\end{tabular}
\end{table*}

\begin{figure}
\includegraphics[width=\linewidth]{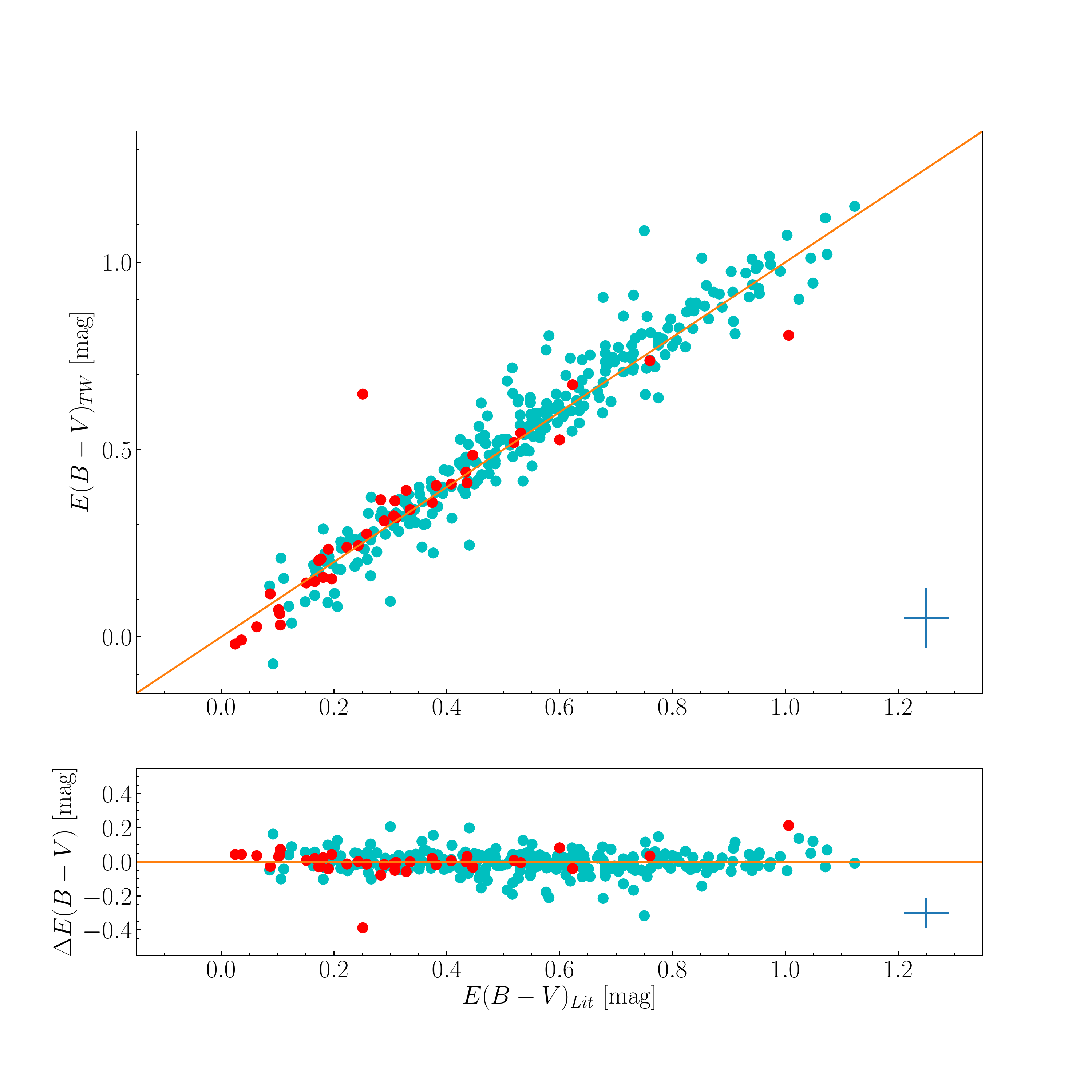}\par
\caption{Comparison between the reddening values estimated from the period-$(V-I)_0$ relation E(B-V)$_{TW}$ and the literature E(B-V)$_{Lit}$ for a sample of 320 GCC (see text for more details). {\sl Top panel} shows the 1:1 diagram; whilst {\sl bottom panel} displays the difference between the two reddening estimates. In both panels, green and red filled circles represent F and FO pulsators, respectively.}
\label{fig:comparisonEBV}
\end{figure}

\begin{figure*}
\begin{multicols}{2}
    \includegraphics[width=\linewidth]{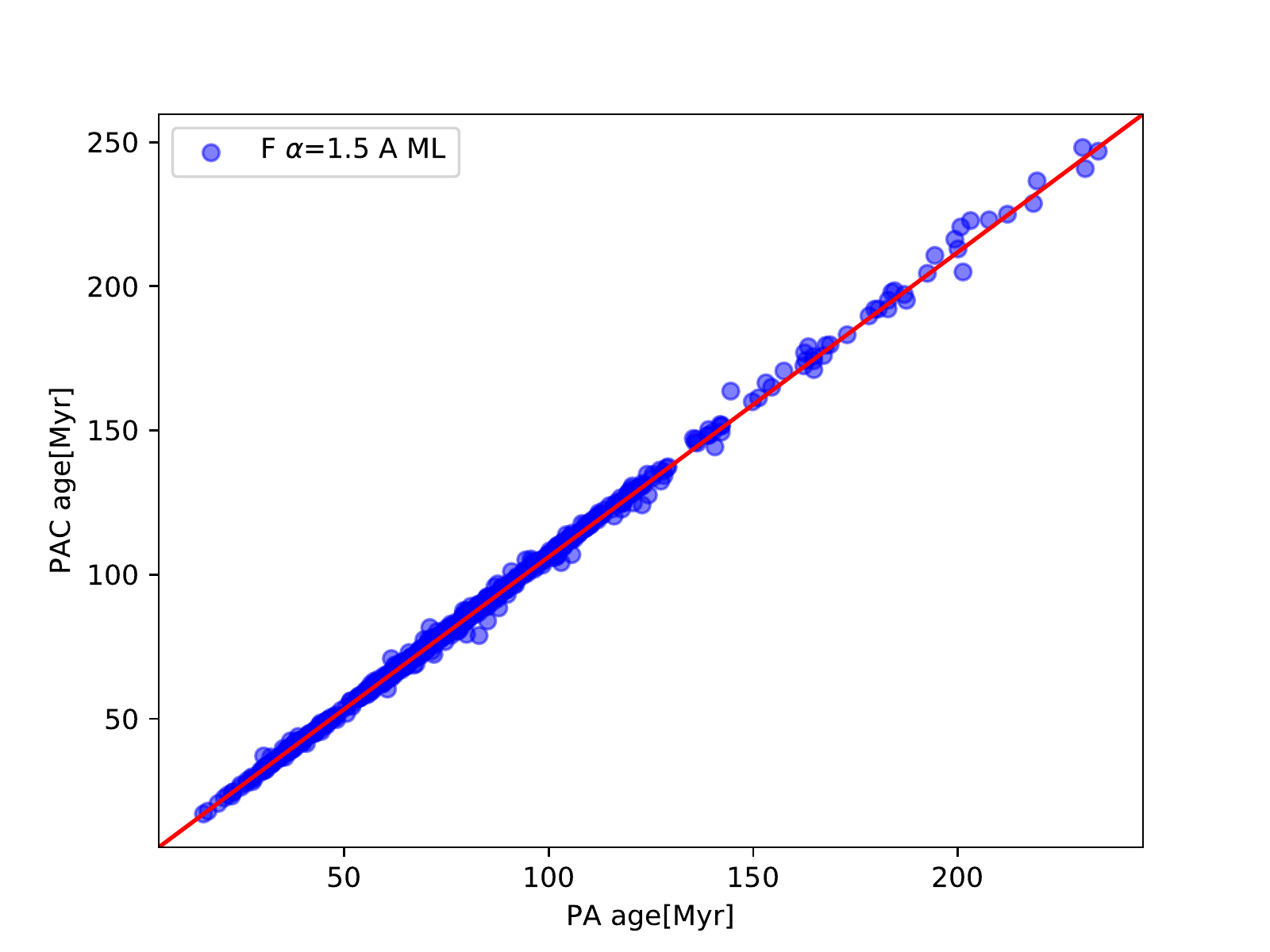}\par 
    \includegraphics[width=\linewidth]{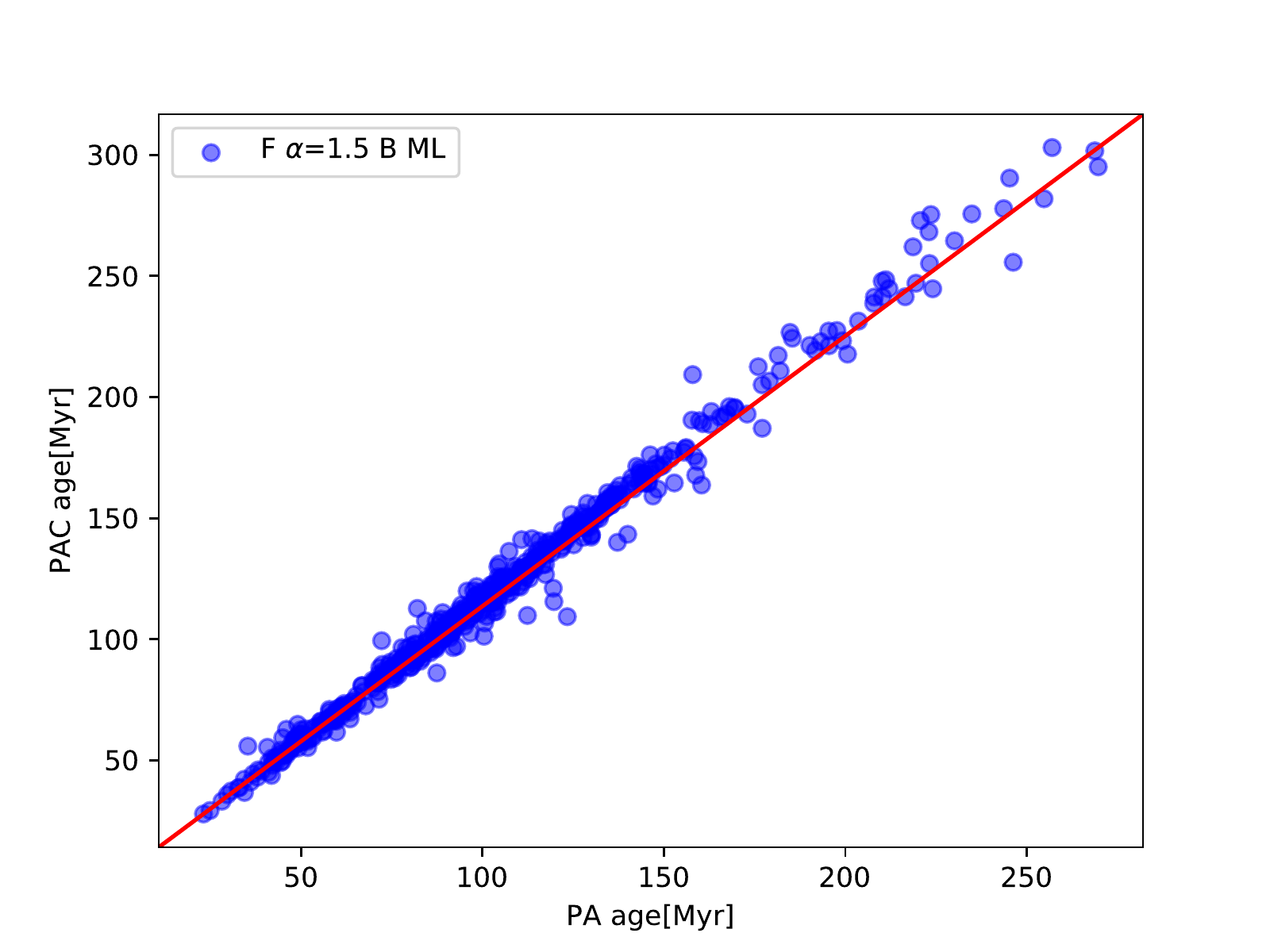}\par 
\end{multicols}
\caption{Comparison between the age estimates obtained by applying the PA and the PAC relations, for the two selected ML cases, (case A in the {\sl left panel} and case B in the {\sl right panel}) to the selected sub-sample of F-mode GCC. In both panels, the solid line represents the 1:1 line.}
\label{fig:PAC_PA_F_AGE_COMP}
\end{figure*}

\begin{figure}
\includegraphics[width=\linewidth]{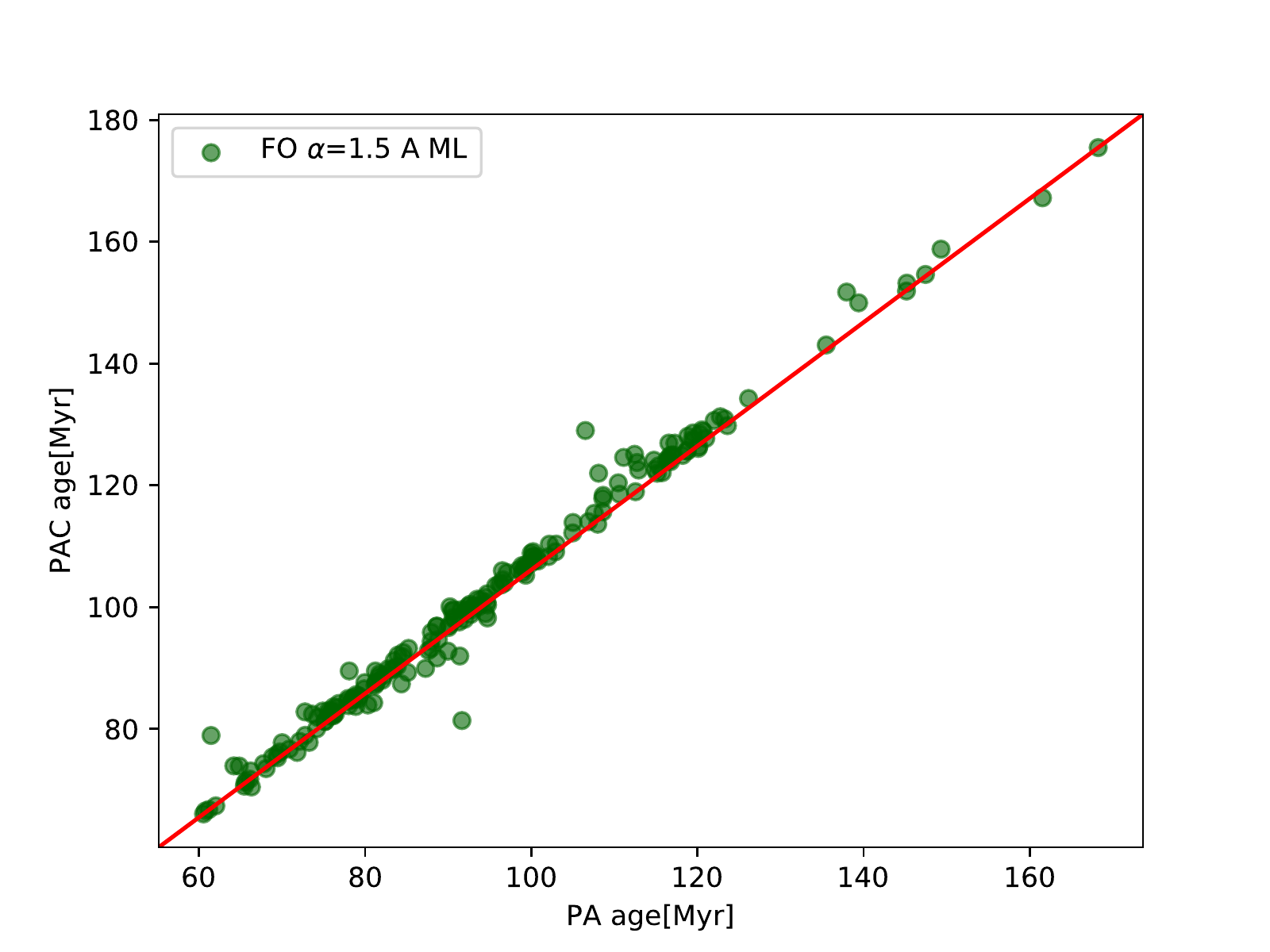}\par
\caption{Comparison between the individual ages obtained by applying the PA and the PAC relations to the selected sub-sample of FO-mode GCC. The solid line represents the 1:1 line.}
\label{fig:PAC_PA_FO_AGE_COMP}
\end{figure}

\section{Individual ages of \textsl{Gaia} DR2 Galactic Classical Cepheids}

In this section we apply the theoretical PA and \textsl{Gaia}-band PAC relations derived above to a sample of GCC published in the \textsl{Gaia} DR2 \citep{Brown2018} and reclassified by \citet{Ripepi2019}. 

The selected sample consists of 568 F-mode and 198 FO-mode pulsators.
Using the period and color values reported by \citet{Ripepi2019} and the PA and PAC relations, we derived the individual ages for each adopted assumption concerning the ML relation. 

Before applying the PAC relation, it is mandatory to obtain dereddened ($G_{BP}-G_{RP}$) colors for the selected GCC sample. For this purpose, for 320 objects in common with our sample we adopted the E(B-V) estimates available in the compilation by \citet{Groenewegen2018}; for all other stars the CC period-(V-I)${_0}$ color relation obtained by Ripepi et al. (in preparation) was used to derive the intrinsic unreddened color. 

The (V-I) color estimate - averaged over the pulsation cycle - was obtained for 89 objects from the catalog of the OGLE Galactic Disk survey \citep{Udalski2018}; while for all other stars\footnote{For 11 multi-mode GCC the PC was not used because the results are less reliable.} the averaged {\it Gaia} DR2 ($G_{BP}-G_{RP}$) color was transformed into the corresponding (V-I) one by using a conversion formula - suitable for the color range spanned by GCC - derived by Ripepi et al. (in preparation).
These E(V-I) values were previously converted into E(B-V) by adopting the relation E(B-V)=1.283$\cdot$E(V-I) \citep{Tammann2003}, and lately into E($G_{BP}-G_{RP}$) by means of Eq. 9-10 in \citet{Ripepi2019}. The average errors on the E(B-V) estimates are $\sim$0.08 mag and $\sim$ 0.11 mag, for the (V-I) color estimated from the OGLE survey and the {\it Gaia} magnitude conversion, respectively.

It is worth checking the level of consistency between the reddening estimates obtained by using the PC relation with those retrieved from the literature for the CC in the compilation of \citet{Groenewegen2018}; figure \ref{fig:comparisonEBV} shows a comparison between the E(B-V) values obtained with the two methods for both F and FO pulsators. The lack of any systematic trend in both the 1:1 (top panel) and the residuals (bottom panel) diagrams, supports the reliability of the approach used for estimating the extinction from the PC relation (Ripepi et al., in preparation). 

We note that the reddening uncertainty affects the age measurements with an uncertainty of $\Delta(log t)\sim$0.009~dex and $\Delta(log t)\sim$ 0.022~dex for F-mode pulsators in cases A and B, respectively. For FO-mode GCC the uncertainty is $\Delta(log t)\sim$0.020~dex. As a consequence, the typical uncertainty on the GCC ages obtained from the PAC relationship, associated with the error on the reddening  estimate, is of the order of 1-2\%.

The GCC sample age estimates obtained by using the PA and PAC relationships are listed in Table \ref{age_F_1.5_A_PA_PAC} to \ref{age_F_1.5_A_FO_PA_PAC}. 

In order to assess the accuracy of our age estimates, in Figures \ref{fig:PAC_PA_F_AGE_COMP} and \ref{fig:PAC_PA_FO_AGE_COMP}, we compare the individual age estimates obtained by applying the PA and the PAC relationships derived in the previous section, for the selected F and FO-mode GCC, respectively. In Figure \ref{fig:PAC_PA_F_AGE_COMP}, the two panels refer to {\sl case A} (left) and {\sl B}  (right). It is worth noting that a good agreement does exist between the age measurements on the whole spanned age range obtained using the two relationships, as proven by the distribution of points with respect to the 1:1 line.

\begin{figure*}
\begin{multicols}{2}
    \includegraphics[width=\linewidth]{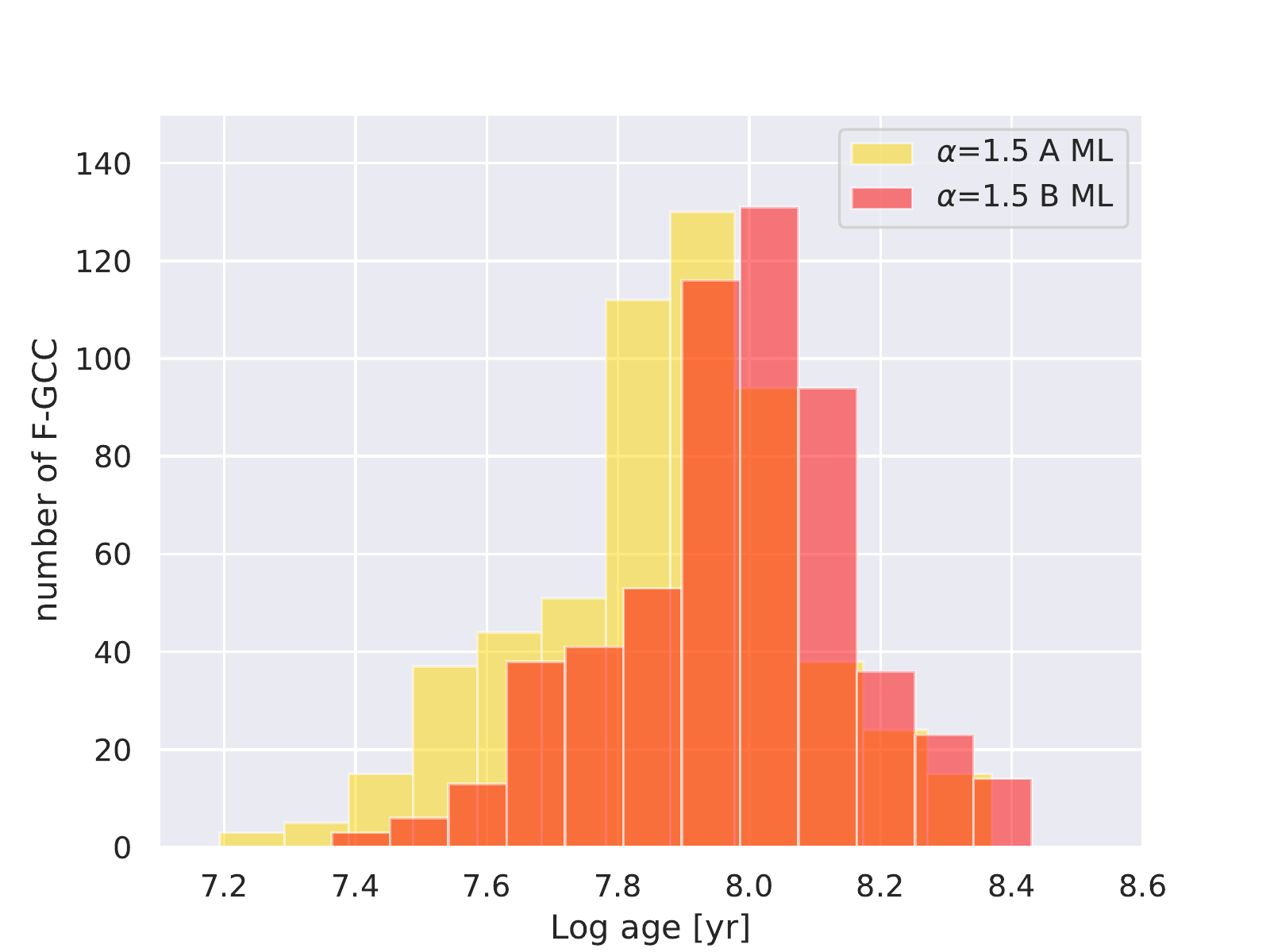}\par 
    \includegraphics[width=\linewidth]{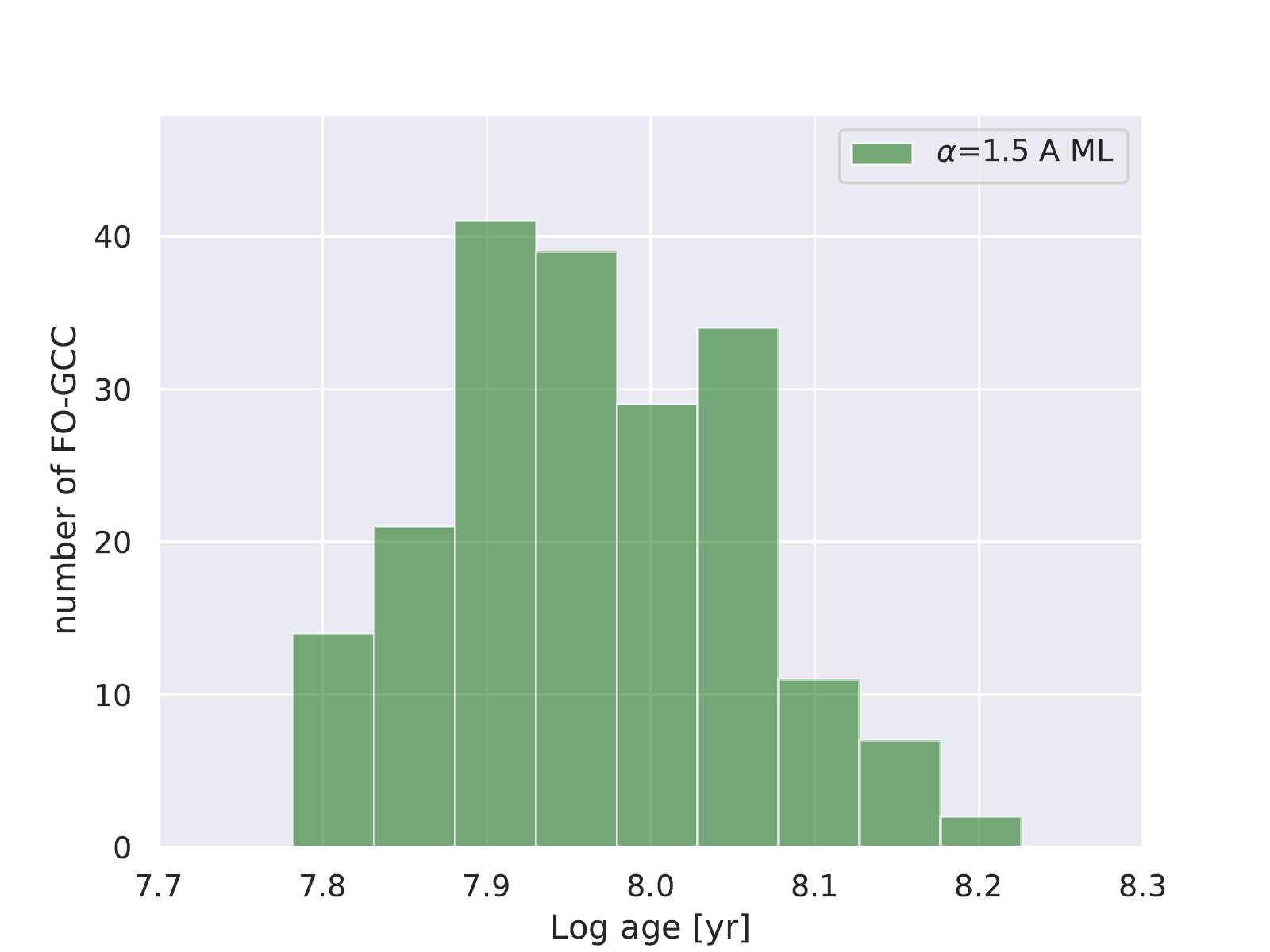}\par 
\end{multicols}
\caption{\label{fig:hist_PAC} The predicted age distribution as obtained by applying the PAC relation to the selected sample of F {\sl (left panel)} and FO-mode {\sl (right panel)} GCC, for the labelled assumptions on the efficiency of superadiabatic convection and ML relation}
\end{figure*}

The histogram of the retrieved age distribution based on the application of the PAC relation, is shown in Figure~\ref{fig:hist_PAC}, for both F (left panel) and FO (right panel) pulsators. Inspection of this figure confirms that the predicted ages depend on the assumed ML relation. Thus, providing systematically older ages for brighter ML relations, in agreement with independent evaluations in the literature \citep[][]{Anderson2016,Bono2005,Senchyna2015}.
For F-mode GCC, the left panel of Figure~\ref{fig:hist_PAC} suggests that the age distribution peaks around $\sim90$~Myr in the assumption of a canonical ML relation ({\sl case A}) and gets older by about $\sim35$~Myr in the non-canonical assumption ({\sl case B}). 
In the case of FO GCC (right panel of Figure~\ref{fig:hist_PAC}), the inferred age distribution is - as expected - concentrated towards older ages with a peak between $(80 - 85)$~Myr.

\subsection{The age map distribution of Gaia DR2 Cepheids}

The possibility of retrieving accurate and reliable age estimates of a large sample of GCC by using the PA and/or the PAC relationships provides the opportunity to perform an age tomography of the Galaxy, and hence to properly trace the Star Formation history of the various portions of the Milky Way, in particular, in the regime of young and intermediate ages.

In order to show the inferred GCC age distribution as a function of the position in the Galactic disk, the selected Gaia DR2 Cepheid sample was represented in the Galactic coordinates with the predicted individual age varying according to the logarithmic color-bar scale and increasing from blue to red. 

\begin{figure*}
\includegraphics[width=\textwidth]{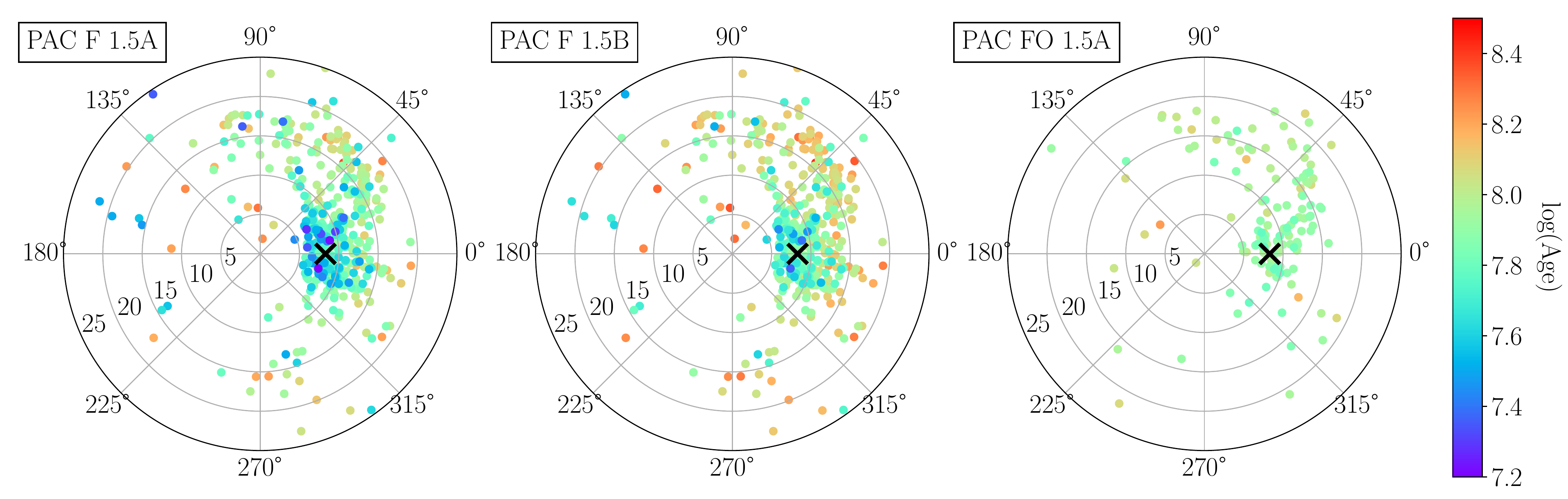}
\caption{Distribution of the selected Gaia DR2 Cepheids on the Galactic plane plotted in polar coordinates. The Galactic center is in the middle. The Galactocentric polar coordinate is 0$\degr$ in the direction of the Sun, whose position is marked by a black cross. Each circle shows Galactocentric distances increasing by 5 kpc, from 0 up to 25 kpc. In each panel the colored circles show the predicted individual ages as obtained by using the PAC relations, according to the logarithmic color-bar axis. The labels on the top-left of each figure indicate the pulsation mode, while 1.5 A or B refer to the PAC obtained by assuming $\alpha_{ml}~$=$1.5$ and canonical and non canonical stellar models, respectively.}
\label{fig:age_map}
\end{figure*}

The maps shown in Figure \ref{fig:age_map} refer to F and FO-mode GCC with ages derived from the PAC relations. From left to right, the panels show the age map for: F-mode GCC by assuming the ML case A, F-mode GCC by assuming the ML case B and FO-mode GCC by assuming the ML case A. In all panels, the pulsation predictions were based on an adopted mixing length equal to $\alpha_{ml}~$=$1.5$.
Inspection of these figures suggests that:
\begin{itemize}
    \item the predicted ages decrease towards the Galactic center, with the oldest Cepheids located at longer Galactocentric distances.
    \item for the same assumption on the ML relation, FO-mode Cepheids are found to be systematically older than the F ones.
\end{itemize}

Although a detailed analysis of this topic is out of the aims of the present investigation, both the age and the spatial distributions of the CC in our sample as a function of their age can be compared with independent studies on the presence of age gradients among the various stellar populations in the Milky Way. For example, \citet{Skowron2019}, applied the PA relation by \citet{Anderson2016} to the OGLE database of GCC, and found an age distribution (see their Figure 3) very similar to our results based on PA and PAC relations obtained from case B models.
Recently, \citet{Bossini2019} computed the age distribution of 269 Galactic open clusters with astrometric and photometric data from Gaia DR2. However, in this case the possibility of realizing a meaningful comparison is hampered by the fact that the open cluster sample adopted by \citet{Bossini2019} spans an age range limited with respect to that of the GCC sample adopted in the present investigation.

\section{Conclusions and Future Developments}

The most recent updated evolutionary predictions for the solar chemical composition based on the BASTI stellar evolution database \citep[][]{hidalgo:18} were combined with the recent nonlinear convective Cepheid models developed by \citetalias{Desomma2020}. This was done to derive updated and accurate PA relations, as well as the first PAC relations in the {\it Gaia} DR2 photometric passbands, for various assumptions about the ML relations and the efficiency of superadiabatic convection in pulsation models.

The coefficients of these relations and hence, the corresponding age predictions, significantly depend on the assumed ML relations with the non canonical ML relation ({\sl case B} in the present work) providing ages older than the canonical ML ({\sl case A}) relation.
For the case of PA relations, in particular, the predicted age difference percentages inferred from {\sl case A} to {\sl case B}, range from $\sim$ 36 \% to $\sim$ 60 \%.
This  trend is confirmed by the behaviour of the PAC relations.

The application of the new PA and {\it Gaia}-band PAC relations to a selected sample \citep{Ripepi2019} of {\it Gaia} DR2 GCC, produces individual ages which systematically increase as the ML relation changes from canonical ({\sl case A}) to non canonical ({\sl case B}) models. When adopting {\sl case A} ML models, the inferred age distribution of F-mode GCC in our sample peaks at around 90~Myr, while for {\sl case B} the peak is shifted upwards by $\sim35$~Myr. This occurrence holds when adopting both the PA and the PAC relations.

On the other hand, the age estimates obtained by means of both the PA and PAC relations are almost insensitive to variations in the mixing length parameter. This result is expected because a variation in the efficiency of superadiabatic convection can modify the pulsation amplitude and the boundary of the IS \citepalias[see e.g.][and references therein]{Desomma2020,Fiorentino2007}, but does not affect at all the  relation between period, color and luminosity; and hence, the relation between period, color and age.

The FO-mode GCC from the sample by \citet{Ripepi2019} are found to be significantly older than the average F pulsator age distribution. This is due to the smaller masses and shorter periods of the FO pulsators.

We have performed a preliminary comparison between the age map distribution obtained in the present work with other similar analyses in literature. In general, we found that older Cepheids are located at a longer Galactocentric distance than younger pulsators, an occurrence in fine agreement with previous results for the age distribution of Milky Way CC obtained by \citet{Skowron2019}.

A more detailed analysis of the inferences that can be obtained from the star formation history of the various portions of our Galaxy at different locations from the study of the age distribution of CC, does need to be supplemented with an accurate investigation of the metallicity distribution of these pulsators. We plan to extend in a future research the present theoretical investigation to different metallicity regimes in order to properly account for the whole metallicity distribution of Galactic Cepheids.

\section*{Acknowledgements}
We thank the anonymous Referee for her/his pertinent and useful comments that significantly improved the content and readability of the manuscript.
We acknowledge Istituto Nazionale di Fisica Nucleare (INFN), Naples section, specific initiative QGSKY. 
This work has made use of data from the European
Space Agency (ESA) mission \textsl{Gaia} (https://www.cosmos.esa.int/gaia),
processed by the \textsl{Gaia} Data Processing and Analysis Consortium (DPAC, https:
//www.cosmos.esa.int/web/gaia/dpac/consortium). Funding for the
DPAC has been provided by national institutions, in particular the institutions
participating in the Gaia Multilateral Agreement. In particular, the Italian participation
in DPAC has been supported by Istituto Nazionale di Astrofisica (INAF)
and the Agenzia Spaziale Italiana (ASI) through grants I/037/08/0, I/058/10/0, 2014-025-R.0, 2014-025-R.1.2015 and 2018-24-HH.0 to INAF (PI M.G. Lattanzi). We acknowledge partial financial support from 'Progetto Premiale' MIUR MITIC (PI B. Garilli). This work has been partially supported by the INAF Main Stream SSH program, 1.05.01.86.28.
SC acknowledges support from Istituto Nazionale di Fisica Nucleare (INFN) (Iniziativa specifica TAsP), grant AYA2013- 42781P from the Ministry of Economy and Competitiveness of Spain, and grant INAF Mainstream (PI: S. Cassisi). This work has made use of the
VizieR database, operated at CDS, Strasbourg, France.

\begin{table*}
\caption{\label{age_F_1.5_A_PA_PAC} Individual ages for the F-mode GCC in our sample as obtained by using both the canonical PA and PAC relations. The column 10 provides information on the reddening estimate: {\sl L} stands for data from literature, while {\sl O} and {\sl G} refer to the cases for which the extinction was estimated by adopting the PC relation, using (V-I) color from the OGLE survey and the conversion to the {\it Gaia} color, respectively (see text for more details). The columns from 11 to 14 list the age and the estimated uncertainty as obtained by alternatively using the PA relation or the PAC one. This table is available in its entirety in machine-readable form.}
\centering
\resizebox{\textwidth}{!}{%
\begin{tabular}{cccccccccccccc}
\hline\hline
\textsl{Gaia} DR2 Source Id & RA[deg] & DEC[deg] & P[d] & $G$[mag] & $G_{BP}$[mag] & $G_{RP}$[mag] & E($G_{BP}$-$G_{RP}$)[mag] & $\sigma$E($G_{BP}$-$G_{RP}$)[mag] & Note & $t_{PA}$[Myr] & $\sigma$ $t_{PA}$[Myr] & $t_{PAC}$[Myr] & $\sigma$ $t_{PAC}$[Myr] \\
\hline
(1)&(2)&(3)&(4)&(5)&(6)&(7)&(8)&(9)&(10)&(11)&(12)&(13)&(14)\\
\hline
3442172745919329664 & 82.35887 & 27.00089 & 3.34920 & 11.57 & 12.21 & 10.78 & 0.72 & 0.12 & G & 105.55 & 20.41 & 98.88 & 18.9 \\
4704080802304630784 & 12.82014 & -68.98430 & 0.99469 & 18.24 & 18.45 & 17.79 & 0.18 & 0.11 & G & 248.10 & 47.99 & 230.53 & 44.06 \\
5521400228203695232 & 123.35129 & -42.27568 & 1.00189 & 16.05 & 16.93 & 15.15 & 1.23 & 0.15 & G & 246.84 & 47.74 & 234.34 & 44.79 \\
6379351625245757568 & 359.99997 & -75.19496 & 1.03777 & 17.68 & 17.98 & 17.36 & 0.02 & 0.11 & G & 240.80 & 46.58 & 231.16 & 44.18 \\
4711142067840447360 & 21.61608 & -64.87535 & 1.06447 & 17.86 & 18.05 & 17.41 & 0.16 & 0.11 & G & 236.54 & 45.75 & 219.39 & 41.93 \\
4639539705975935360 & 50.85364 & -74.96868 & 1.11640 & 17.83 & 18.12 & 17.50 & 0.02 & 0.11 & G & 228.74 & 44.24 & 218.50 & 41.76 \\
4690768156035338624 & 18.95551 & -70.54573 & 1.14292 & 17.53 & 17.82 & 17.09 & 0.18 & 0.11 & G & 224.99 & 43.52 & 212.14 & 40.54 \\
4703965697180635136 & 9.31520 & -67.04117 & 1.15729 & 17.89 & 18.12 & 17.43 & 0.17 & 0.11 & G & 223.02 & 43.14 & 207.66 & 39.69 \\
4666616485480394368 & 60.40851 & -69.47789 & 1.15905 & 16.29 & 16.45 & 15.90 & 0.12 & 0.08 & O & 222.78 & 43.09 & 203.12 & 38.82 \\
4691081375113472256 & 20.83822 & -70.57775 & 1.17564 & 17.33 & 17.45 & 17.02 & 0.0 & 0.11 & G & 220.56 & 42.66 & 200.73 & 38.36 \\
4637614151878607232 & 20.74650 & -76.24745 & 1.20857 & 17.80 & 17.95 & 17.36 & 0.1 & 0.11 & G & 216.31 & 41.84 & 199.28 & 38.09 \\
5834840099568788864 & 241.98124 & -58.78225 & 1.23577 & 13.71 & 14.05 & 13.19 & 0.31 & 0.11 & G & 212.95 & 41.19 & 200.06 & 38.23 \\
4636112425153243904 & 16.19343 & -76.86050 & 1.25457 & 17.34 & 17.55 & 16.84 & 0.21 & 0.11 & G & 210.70 & 40.75 & 194.40 & 37.15 \\
4690721839108141568 & 19.01385 & -70.91305 & 1.30481 & 17.25 & 17.73 & 17.00 & -0.0 & 0.11 & G & 204.95 & 39.64 & 201.29 & 38.47 \\
4698739817197286912 & 23.24870 & -66.49795 & 1.30920 & 17.05 & 17.30 & 16.67 & 0.05 & 0.11 & G & 204.47 & 39.55 & 192.58 & 36.8 \\
5298606801235832064 & 140.10591 & -62.22648 & 1.36601 & 18.36 & 18.60 & 17.89 & 0.18 & 0.11 & G & 198.45 & 38.38 & 184.56 & 35.27 \\
... & ... & ... & ... & ... & ... & ... & ... & ... & ... & ... & ... & ... & ... \\
\hline\hline
\end{tabular}}
\end{table*}

\begin{table*}
\caption{\label{age_F_1.5_B_PA_PAC} As Table~\ref{age_F_1.5_A_PA_PAC}, but in this case the PA and PAC relations for the non canonical stellar models have been adopted. This table is available in its entirety in machine-readable form.}
\centering
\resizebox{\textwidth}{!}{%
\begin{tabular}{cccccccccccccc}
\hline\hline
\textsl{Gaia} DR2 Source Id & RA[deg] & DEC[deg] & P[d] & $G$[mag] & $G_{BP}$[mag] & $G_{RP}$[mag] & E($G_{BP}$-$G_{RP}$)[mag] & $\sigma$E($G_{BP}$-$G_{RP}$)[mag] & Note & $t_{PA}$[Myr] & $\sigma$ $t_{PA}$[Myr] & $t_{PAC}$[Myr] & $\sigma$ $t_{PAC}$[Myr] \\
\hline
(1)&(2)&(3)&(4)&(5)&(6)&(7)&(8)&(9)&(10)&(11)&(12)&(13)&(14)\\
\hline
3442172745919329664 & 82.35887 & 27.00089 & 3.34920 & 11.57 & 12.21 & 10.78 & 0.72 & 0.12 & G & 141.71 & 26.10 & 122.49 & 21.72 \\
4704080802304630784 & 12.82014 & -68.98430 & 0.99469 & 18.24 & 18.45 & 17.79 & 0.18 & 0.11 & G & 303.00 & 55.82 & 257.01 & 45.57 \\
5521400228203695232 & 123.35129 & -42.27568 & 1.00189 & 16.05 & 16.93 & 15.15 & 1.23 & 0.15 & G & 301.64 & 55.56 & 268.78 & 47.65 \\
6379351625245757568 & 359.99997 & -75.19496 & 1.03777 & 17.68 & 17.98 & 17.36 & 0.02 & 0.11 & G & 295.07 & 54.35 & 269.73 & 47.82 \\
4711142067840447360 & 21.61608 & -64.87535 & 1.06447 & 17.86 & 18.05 & 17.41 & 0.16 & 0.11 & G & 290.41 & 53.50 & 245.31 & 43.49 \\
4639539705975935360 & 50.85364 & -74.96868 & 1.11640 & 17.83 & 18.12 & 17.50 & 0.02 & 0.11 & G & 281.88 & 51.92 & 254.78 & 45.17 \\
4690768156035338624 & 18.95551 & -70.54573 & 1.14292 & 17.53 & 17.82 & 17.09 & 0.18 & 0.11 & G & 277.77 & 51.17 & 243.67 & 43.20 \\
4703965697180635136 & 9.31520 & -67.04117 & 1.15729 & 17.89 & 18.12 & 17.43 & 0.17 & 0.11 & G & 275.60 & 50.77 & 234.88 & 41.64 \\
4666616485480394368 & 60.40851 & -69.47789 & 1.15905 & 16.29 & 16.45 & 15.90 & 0.12 & 0.08 & O & 275.34 & 50.72 & 223.60 & 39.64 \\
4691081375113472256 & 20.83822 & -70.57775 & 1.17564 & 17.33 & 17.45 & 17.02 & 0.0 & 0.11 & G & 272.90 & 50.27 & 220.70 & 39.13 \\
4637614151878607232 & 20.74650 & -76.24745 & 1.20857 & 17.80 & 17.95 & 17.36 & 0.1 & 0.11 & G & 268.22 & 49.41 & 223.07 & 39.55 \\
5834840099568788864 & 241.98124 & -58.78225 & 1.23577 & 13.71 & 14.05 & 13.19 & 0.31 & 0.11 & G & 264.51 & 48.73 & 230.11 & 40.80 \\
4636112425153243904 & 16.19343 & -76.86050 & 1.25457 & 17.34 & 17.55 & 16.84 & 0.21 & 0.11 & G & 262.02 & 48.27 & 218.67 & 38.77 \\
4690721839108141568 & 19.01385 & -70.91305 & 1.30481 & 17.25 & 17.73 & 17.00 & -0.0 & 0.11 & G & 255.66 & 47.09 & 246.30 & 43.67 \\
4698739817197286912 & 23.24870 & -66.49795 & 1.30920 & 17.05 & 17.30 & 16.67 & 0.05 & 0.11 & G & 255.13 & 47.00 & 223.23 & 39.58 \\
5298606801235832064 & 140.10591 & -62.22648 & 1.36601 & 18.36 & 18.60 & 17.89 & 0.18 & 0.11 & G & 248.43 & 45.76 & 211.15 & 37.44 \\
... & ... & ... & ... & ... & ... & ... & ... & ... & ... & ... & ... & ... & ... \\
\hline\hline
\end{tabular}}
\end{table*}

\begin{table*}
\caption{\label{age_F_1.5_A_FO_PA_PAC} As Table~\ref{age_F_1.5_A_PA_PAC}, but for the FO-mode GCC in the selected sample. This table is available in its entirety in machine-readable form.}
\centering
\resizebox{\textwidth}{!}{%
\begin{tabular}{cccccccccccccc}
\hline\hline
\textsl{Gaia} DR2 Source Id & RA[deg] & DEC[deg] & P[d] & $G$[mag] & $G_{BP}$[mag] & $G_{RP}$[mag] & E($G_{BP}$-$G_{RP}$)[mag] & $\sigma$E($G_{BP}$-$G_{RP}$)[mag] & Note & $t_{PA}$[Myr] & $\sigma$ $t_{PA}$[Myr] & $t_{PAC}$[Myr] & $\sigma$ $t_{PAC}$[Myr] \\
\hline
(1)&(2)&(3)&(4)&(5)&(6)&(7)&(8)&(9)&(10)&(11)&(12)&(13)&(14)\\
\hline
5958267083020200448 & 263.74736 & -44.83491 & 0.48582 & 15.31 & 15.67 & 14.78 & 0.48 & 0.11 & G & 175.45 & 21.01 & 168.26 & 17.82 \\
4652801401061740800 & 72.89387 & -73.52919 & 0.54856 & 17.88 & 18.20 & 17.42 & 0.32 & 0.11 & G & 167.21 & 20.02 & 161.55 & 17.11 \\
4649684869708905600 & 78.33946 & -73.18860 & 0.62515 & 17.60 & 17.85 & 17.14 & 0.28 & 0.11 & G & 158.78 & 19.01 & 149.33 & 15.82 \\
4757521942202075904 & 84.00743 & -62.92647 & 0.66840 & 17.23 & 17.49 & 16.81 & 0.21 & 0.11 & G & 154.63 & 18.51 & 147.46 & 15.62 \\
4648894973678313600 & 78.74744 & -76.16060 & 0.68428 & 17.58 & 17.85 & 17.13 & 0.26 & 0.11 & G & 153.20 & 18.34 & 145.22 & 15.38 \\
426097765508018560 & 13.85755 & 59.72753 & 0.69910 & 13.67 & 14.18 & 12.99 & 0.72 & 0.12 & G & 151.90 & 18.19 & 145.17 & 15.38 \\
5255955821811454592 & 150.19748 & -61.57770 & 0.70107 & 14.74 & 15.09 & 14.05 & 0.65 & 0.12 & G & 151.73 & 18.17 & 137.97 & 14.61 \\
6380124062227694208 & 357.83150 & -72.69153 & 0.72213 & 17.86 & 18.07 & 17.44 & 0.19 & 0.11 & G & 149.96 & 17.96 & 139.41 & 14.77 \\
4648520761770393088 & 85.76598 & -75.18605 & 0.81359 & 17.25 & 17.54 & 16.80 & 0.25 & 0.11 & G & 143.05 & 17.13 & 135.51 & 14.35 \\
4652889258915055488 & 72.48404 & -73.00018 & 0.95488 & 17.24 & 17.54 & 16.77 & 0.27 & 0.11 & G & 134.26 & 16.08 & 126.16 & 13.36 \\
5542032357742851712 & 125.44982 & -37.46816 & 1.01087 & 16.30 & 17.08 & 15.38 & 1.19 & 0.12 & O & 131.26 & 15.72 & 122.76 & 13.00 \\
4652312599416423040 & 68.88711 & -74.56297 & 1.01787 & 17.03 & 17.30 & 16.59 & 0.18 & 0.11 & G & 130.90 & 15.67 & 123.28 & 13.06 \\
4649732762885142272 & 77.24295 & -73.85941 & 1.02268 & 17.06 & 17.35 & 16.58 & 0.26 & 0.11 & G & 130.66 & 15.64 & 122.03 & 12.92 \\
5860021737714232576 & 190.95445 & -65.86875 & 1.03985 & 15.64 & 16.14 & 14.78 & 0.82 & 0.1 & O & 129.80 & 15.54 & 123.61 & 13.09 \\
4690203483799384832 & 12.47494 & -70.52511 & 1.05413 & 16.86 & 17.06 & 16.51 & 0.03 & 0.11 & G & 129.10 & 15.46 & 120.55 & 12.77 \\
5879216457587855872 & 220.33866 & -58.95261 & 1.05620 & 17.35 & 19.40 & 15.97 & 3.13 & 0.28 & G & 129.00 & 15.45 & 106.51 & 11.28 \\
... & ... & ... & ... & ... & ... & ... & ... & ... & ... & ... & ... & ... & ... \\
\hline\hline
\end{tabular}}
\end{table*}

\clearpage

Data availability statement:The data underlying this article are available in the article and in its online supplementary material.



\bibliographystyle{mnras}
\bibliography{giu} 


\bsp	
\label{lastpage}
\end{document}